\documentclass{aa}  

\usepackage{graphicx}
\usepackage{txfonts}
\usepackage{natbib}

\usepackage{color}

\newcommand{\Msun}{M$_{\odot}$}
\newcommand{\Msunyr}{M$_{\odot}$\,yr$^{-1}$}

\newcommand{\kms}{km\,s$^{-1}$}

\begin{document}
   \title{The detached dust shells around the carbon AGB stars \\R~Scl and V644~Sco}


   \author{M. Maercker \inst{1,2}
   \and
          S. Ramstedt \inst{3}
         M. L. Leal-Ferreira \inst{2}
          \and
          G. Olofsson \inst{4}
    	\and
          H.G. Floren \inst{4}
          }

     \institute{Onsala Space Observatory, Dept. of Radio and Space Science, Chalmers University of Technology, SE-43992 Onsala, Sweden\\
   \email{maercker@chalmers.se}
   \and
   Argelander Institut f\"ur Astronomie, University of Bonn, Auf dem H\"ugel 71, D-53121 Bonn, Germany
   \and
   Department of Physics and Astronomy, Uppsala University, Box 516, 751 20 Uppsala, Sweden
   \and
   Department of Astronomy, Stockholm University, AlbaNova University Center, 106 91 Stockholm, Sweden
             }

   \date{Submitted 1 September 2014; accepted for publication in A\&A}

  \abstract
  {The morphology of the circumstellar envelopes (CSE) around Asymptotic Giant Branch (AGB) stars gives information on the mass-loss process from the star, its evolution and wind, and the effect of binary interaction. However, determining the distribution of dust in the circumstellar envelopes is difficult. Observations of polarised, dust-scattered stellar light in the optical have produced images with high-spatial resolution of the envelopes around evolved stars. In particular for sources with detached shells this method has proven extremely successful. Detached shells are believed to be created during a thermal pulse, and constrain the time scales and physical properties of one of the main drivers of late stellar evolution.}
   {We aim at determining the morphology of the detached shells around the carbon AGB stars R~Scl and V644~Sco. In particular, we attempt to constrain the radii and widths of the detached dust shells around the stars and compare these to observations of the detached gas shells.}
   {We observe the polarised, dust-scattered stellar light around the carbon AGB stars R~Scl and V644~Sco using the PolCor instrument mounted on the ESO 3.6m telescope. Observations were done with a coronographic mask to block out the direct stellar light. The polarised images clearly show the detached shells around R~Scl and V644~Sco. Using a dust radiative transfer code to model the dust-scattered polarised light we constrain the radii and widths of the shells.}
   {We determine radii of 19\farcs5 and 9\farcs4 for the detached dust shells around R~Scl and V644~Sco, respectively. Both shells have an overall spherical symmetry and widths of $\approx$2\arcsec. For R~Scl we can compare the observed dust emission directly with high spatial-resolution maps of CO($3-2$) emission from the shell observed with ALMA. We find that the dust and gas coincide almost exactly, indicating a common evolution. The data presented here for R~Scl are the most detailed observations of the entire dusty detached shell to date. For V644~Sco these are the first direct measurements of the detached shell. Also here we find that the dust most likely coincides with the gas shell.}
   {The observations are consistent with a scenario where the detached shells are created during a thermal pulse. The determined radii and widths will constrain hydrodynamical models describing the pre-pulse mass loss, the thermal pulse, and post-pulse evolution of the star.}

\titlerunning{The detached dust shells around R~Scl and V644~Sco}
   \maketitle

\section{Introduction}
\label{s:intro}

During the late stages of stellar evolution, asymptotic giant branch (AGB) stars build up large circumstellar envelopes (CSEs) of dust and gas. The CSEs are a direct consequence of the mass loss the stars experience on the AGB, carrying material processed inside the star to the interstellar medium (ISM). CSEs are generally assumed to be formed through an isotropic mass loss from the stellar surface, building up a spherically symmetric envelope \citep[e.g.,][]{castrocarrizoetal2010}. Through their relatively simple geometry, the CSEs around AGB stars are excellent astrochemical and astrophysical laboratories. However, variations in the mass-loss rate over time, and any directional dependence of the mass loss will influence the shape of the CSE. Recent images of thermal dust emission in the far-infrared taken with the PACS instrument aboard the Herschel Space Telescope show a variety of envelope shapes on very large scales \citep{coxetal2012}, including arc-like structures \citep{decinetal2012}. Spiral-shaped CSEs have been found around a handful of sources, e.g., AFGL 3068 \citep{mauronco2006}, CIT~6 \citep{kimetal2013}, o Ceti \citep{mayeretal2011}. The most detailed observations so far were taken with the Atacama Large Millimeter/submillimeter Array (ALMA) of the CSE around the carbon star R~Scl  in molecular line emission \citep[][hereafter M2012]{maerckeretal2012}. They show a spiral structure extending from the detached shell inwards to the central star and present-day mass loss, indicating the presence of a binary companion shaping the wind. The asymmetries observed in the CSEs around AGB stars may be due to several processes. Non-isotropic mass loss, binary interaction, and outflows can shape the envelopes around AGB stars into asymmetric shapes \citep{mastrodemosco1999, mauronco2006, kimco2012}. Temporal variations in the mass loss will create rings and/or arc-like structures in form of density enhancements in the envelope. An example of the most extreme cases of this can be seen in the detached shells around a number of carbon AGB stars, believed to be caused by the increase in mass-loss rate and expansion velocity during a thermal pulse \citep{olofssonetal1990,olofssonetal1993a,olofssonetal1996,steffenetal1998,steffenco2000,schoieretal2005,mattssonetal2007,maerckeretal2010,olofssonetal2010,maerckeretal2012}. 

The duration of a thermal pulse is on the order of a few hundred years, with timescales between pulses being on the order of several 10000 years \citep{karakasco2007}. The chances of observing a pulse directly are extremely small. Currently less than $\approx$10 detached shells of gas and dust formed during a thermal pulse are known around carbon stars. Considering the typical time-scales between thermal pulses and the lifetime of a detached shell of CO gas due to photodissociation, it is not likely that more sources of this kind will be found \citep{olofssonetal1990}. Detailed observations of these detached shell sources may hence be the only way for us to effectively constrain the evolution of the star throughout the thermal pulse cycle. The spiral structure observed around R~Scl in the ALMA observations allowed to measure the evolution of the mass-loss rate and expansion velocity since the last thermal pulse (M2012), and is the only case where direct determination has been possible.

Direct imaging of the envelopes is difficult, in particular of the extended, dusty CSE. Thermal dust emission in the far infrared from detached shells has been observed by the AKARI instrument~\citep{izumiuraetal2011} and the PACS instrument~\citep{kerschbaumetal2010,mecinaetal2014}. PACS in particular delivered comparatively high-resolution images of a large number of AGB CSEs~\citep{coxetal2012}.

Observations of dust scattered stellar light have proven to be an effective way to obtain high-resolution images of the dusty CSEs. Hubble Space Telescope images of R~Scl give the highest-resolution images of the detached shell of dust around this object to date \citep[][hereafter O2010]{olofssonetal2010}, although only approximately one-third of the shell could be imaged. Images of polarised, dust-scattered stellar light add a powerful tool to investigate the distribution of the dust in the plane of the sky. This technique was used to successfully image the detached shells around R~Scl and U~Ant (\citealt{delgadoetal2003a}, hereafter GD2003; \citealt{maerckeretal2010}). The detached shells around DR~Ser and U~Cam, as well as the circumstellar environment around the S-type AGB star W~Aql, were imaged in polarised, dust-scattered stellar light using the PolCor instrument mounted on the Nordic Optical Telescope~\citep{ramstedtetal2011}. This was the first direct image of the detached shell around the carbon star DR~Ser. Until then the shell had only been inferred indirectly through the line profiles in molecular line observations \citep[][hereafter S2005]{schoieretal2005}. The polarised, dust-scattered light observations allowed to directly measure the radius and thickness of the shell around DR~Ser.

We now present new observations with the PolCor instrument mounted on the ESO 3.6m telescope in La Silla, Chile, of the dusty CSEs around two detached-shell AGB stars. The instrument provides images in polarised light combined with a coronographic mask. This allows us to block the direct, stellar light, making the faint, dust-scattered stellar light in the circumstellar medium visible. In Sect.~\ref{s:obs} we present the observations, source properties, and data treatment, Sect.~\ref{s:radtransf} gives an overview of the radiative transfer modelling, and Sects.~\ref{s:results} to~\ref{s:conclusions} present and discuss the results, and give our conclusions.

\section{Observations and data reduction}
\label{s:obs}

The observations were taken with the PolCor instrument mounted on the ESO 3.6m telescope in La Silla, Chile, during 5 nights between October 18 and October 28 2011. We here present the observations of the carbon AGB stars R~Scl taken on October 19 and V644~Sco taken on October 24, 26, and 27. They were part of a sample of 14 evolved stars to study the morphology of the dusty circumstellar environment around AGB stars. Observations of these sources were taken in the V and R bands centred at wavelengths of 0.55\,$\mu$m and 0.64\,$\mu$m, respectively. The pixel scale in the images is 0\farcs114 per pixel. Typical values of the seeing were $\approx$1\farcs3. All images were flat-fielded and dark-subtracted. PolCor reads out images at rates of 0.04 \,s and 0.1\,s, allowing us to use the \emph{lucky imaging} technique to shift-and-add the individual exposures, effectively improving the seeing in the image. Typically between 10-20\% of the images were rejected in the lucky imaging, resulting in a seeing of $\approx$0\farcs9. In addition, the instrument takes images at four different polarisation angles (0, 45, 90, and 135 degrees) to determine the full set of Stokes parameters \emph{I}, \emph{U}, and \emph{Q}, as well as the polarisation angle. The instrument is equipped with coronographic masks of varying sizes and densities to block the direct stellar light, making the faint scattered light visible. For the observations presented here a mask with a diameter of 3\arcsec\, was used. In order to more easily place the star in the centre of the mask, a mask that reduced the direct stellar light by 99\% was used. Total on-source integration times (before image rejection in the lucky imaging) are listed in Table~\ref{t:observations}. The three nights of observations for V644~Sco were averaged. The PolCor instrument is described in detail in \cite{ramstedtetal2011}.


\begin{table}
\caption{PolCor observations with the 3.6m telescope at La Silla observatory in October 2011.}
\label{t:observations}
\centering
\begin{tabular}{l c c c c}
\hline\hline
Source	& Date	& band		& mask		& t$_{tot}$\\
		&		& 			& ["] 			& [s] \\
\hline
R~Scl	& 19		& V			& 3			& 120\\
		& 19		& R 			& 3			& 120\\
V644~Sco& 24/26/27& V 			& 3			& 750\\
		& 24/26/27& R			& 3			& 1500\\
\hline\hline
\end{tabular}
\end{table}

\subsection{Observed Sources}
\label{s:sources}

\subsubsection{R~Scl}
\label{s:sourcerscl}

R~Scl is a carbon-rich, semi-regular pulsating AGB star with a pulsation period of 370 days. The luminosity is estimated to 4300 L$_{\odot}$ using a period-luminosity relation \citep{knappetal2003}, giving a distance of 290 pc. The parallax gives a distance of 266 pc. The detached shell around R~Scl was first observed by \cite{olofssonetal1990} in CO emission with the single-dish telescope SEST (Swedish-European Submillimeter Telescope). The gas has since been imaged in scattered stellar light in the resonance lines of Na and K \citep[][hereafter GD2001]{delgadoetal2001}, and in high-spatial resolution interferometer maps in CO emission by ALMA (M2012). The dusty detached shell was observed in polarised, dust-scattered stellar light (GD2003), in dust-scattered stellar light with the Hubble Space Telescope (O2010), and in thermal dust emission with the PACS instrument aboard Herschel \citep{coxetal2012}. All observations show a detached shell at 19\arcsec$-$20\arcsec. The expansion velocity of the shell is measured to 14.5\,\kms (M2012), and the present-day expansion velocity to 10.5\,\kms\,~(S2005). The present-day mass-loss rate is estimated to be $3\times10^{-7}$\,\Msunyr\,(S2005), while M2012 determine a mass-loss rate during the creation of the shell of $\approx10^{-5}$\,\Msunyr. The total estimated dust mass in the shell is very uncertain, and ranges between $(2-3)\times10^{-6}$\,\Msun\,(GD2003; O2010) to $3\times10^{-5}$\,\Msun\,(S2005). The gas mass is estimated to be $2.5\times10^{-3}$\,\Msun\,(S2005).

\subsubsection{V644~Sco}
\label{s:sourcerscl}

V644 Sco is an irregularly pulsating carbon AGB star. Its distance is estimated to be 700 pc, based on SED modelling and assuming a luminosity of 4000 L$_{\odot}$ (S2005). V644~Sco was first observed in CO emission using SEST by \cite{olofssonetal1996}, where a detached gas shell at less than 7\arcsec was found. The shells of dust and gas were modelled by S2005, giving estimates of the dust and gas masses of $1.4\times10^{-4}$\,\Msun\, and $2.5\times10^{-3}$\,\Msun, respectively. They derive a radius for the detached gas shell of 10\farcs5.  Previous to this publication, the detached shell had not been observed directly.

\subsection{Polarised dust-scattered stellar light}
\label{s:dustscatter}

Stellar light is unpolarised unless it is intercepted by a medium and scattered. For dust grains the scattering of stellar light occurs in all directions (however, not necessarily isotropically), while the highest degree of polarisation is reached if the angle between the incoming light and the outgoing scattered light is at 90 degrees \citep[e.g.,][]{zubkoco2000}. For stellar light scattered by a dusty CSE surrounding the star, the polarised light hence shows the distribution of the dust in the plane of the sky. Observations at four different polarisation angles can be used to derive the full set of Stokes parameters \emph{I}, \emph{U} and \emph{Q}, as well as the polarised intensity, \emph{P}, and the polarisation degree, $p$~\citep[see, e.g., ][]{maerckeretal2010,ramstedtetal2011}.

We assume that interstellar space and Earth's atmosphere have a negligible effect on the detected light, and that the telescope does not introduce a significant degree of polarisation \citep[e.g., GD2003;][]{ramstedtetal2011}. The measured images then only contain contributions from the star and from the CSE:

\begin{equation}
I_m=I^*+I_{CSE},
\end{equation}
\begin{equation}
U_m=U^*+U_{CSE},
\end{equation}
\begin{equation}
Q_m=Q^*+U_{CSE}
\end{equation}

\begin{equation}
\label{e:P}
P_m=\sqrt{U_m^2+Q_m^2}.
\end{equation}

The direct stellar light is unpolarised, and the measured Stokes U and Q parameters, and hence also the polarised intensity, are equal to the contribution from the circumstellar medium only. Any residual polarisation added by the direct stellar light is generally behind and close to the edges of the mask where interpretation of the data already is difficult, and hence does not affect the analysis significantly.

\subsection{PSF subtraction}
\label{s:psfsub}

In order to correct for the direct, stellar light in the total intensity images, template stars of similar magnitude, but without any (known) circumstellar emission were observed in the same setup as the science targets. The direct stellar contribution to the total intensity can then be subtracted by scaling the brightness of the template star to the science target~\cite[see also GD2001;][]{maerckeretal2010,ramstedtetal2011}.

For several reasons the template subtraction is more complicated than anticipated: 1) the shape of the psf depends on the position of the star behind the mask, 2) the scaling of the template star to the science target is not trivial, as the CSE may contribute to the total intensity all the way to the edge of the mask, and 3) the shape of the psf can differ between science target and template star. Incorrect subtraction of the template star may result in significant artefacts in the final images. The observations of R~Scl and V644~Sco were part of a sample of 14 sources to observe the dusty circumstellar structure around AGB stars. As such, the time spent on each object was limited, making a good psf subtraction difficult.

Not subtracting a template psf leads to a lower limit when calculating the polarisation degree of the circumstellar envelope from Equation. For V644~Sco template subtraction was not possible, and in all cases we present the polarisation degree as a lower limit. Due to the larger angular size of the detached shell around R~Scl, template subtraction was somewhat simpler. In this case, in addition to calculating the lower limit, we also scaled the observed template star to the total intensity images and subtracted the psf. 

Observations of flux template stars in principle allow to absolute flux-calibrate the images. However, the focus of this study is the morphological information contained in the polarised light, and absolute flux calibration is therefore not necessary. 

\section{Dust radiative transfer analysis of scattered stellar light}
\label{s:radtransf}

We used the dust radiative transfer code \emph{radmc-3d} \citep{dullemond2012} in order to model the polarised, circumstellar emission towards R~Scl and V644~Sco. Radmc-3d allows to derive the full set of Stokes vectors for stellar light scattered by dust grains of any type of dust. We used amorphous carbon grains with optical constants from \cite{suh2000} with a grain density of 2 g/cm$^3$ and a grain radius of 0.1\,$\mu$m. 

We modeled the polarised, stellar scattered light assuming a constant present-day mass-loss rate expanding at a constant velocity, giving a r$^{-2}$ density profile. The detached shell is assumed to follow a gaussian density profile in the radial direction with a radius $R$ and a full width at half maximum (FWHM) of $\Delta R$. The parameters for the present-day mass-loss rate and detached shell are taken from S2005. Due to large uncertainties in the dust properties (composition, size, shape), a detailed modelling of the dust to determine new dust masses and dust mass-loss-rates will not improve on previous results, and we instead concentrate on the detailed morphological information of the detached shells.

\section{Results}
\label{s:results}
\begin{figure*}
\centering
\includegraphics[width=4.8cm]{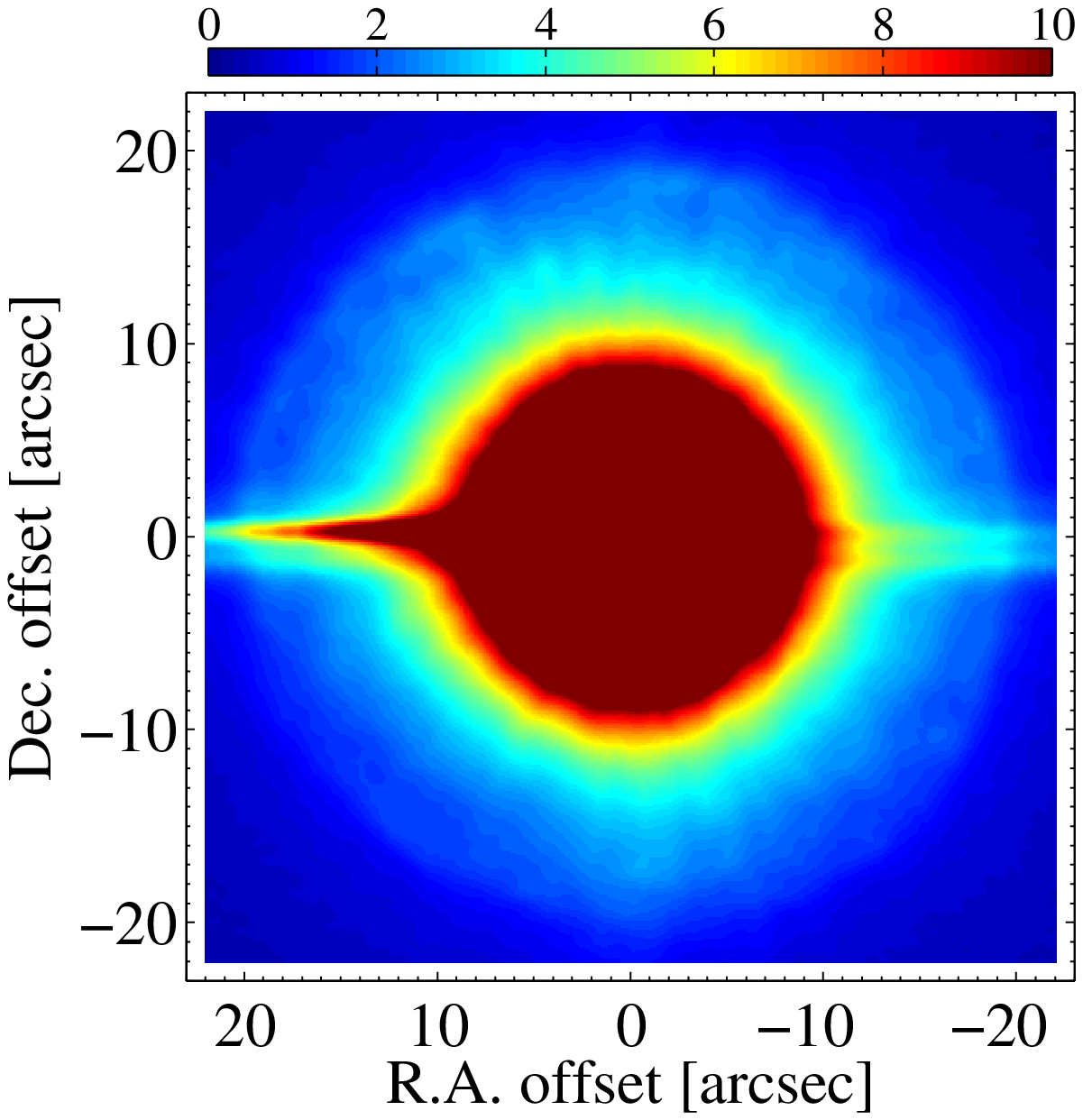}
\includegraphics[width=4.425cm]{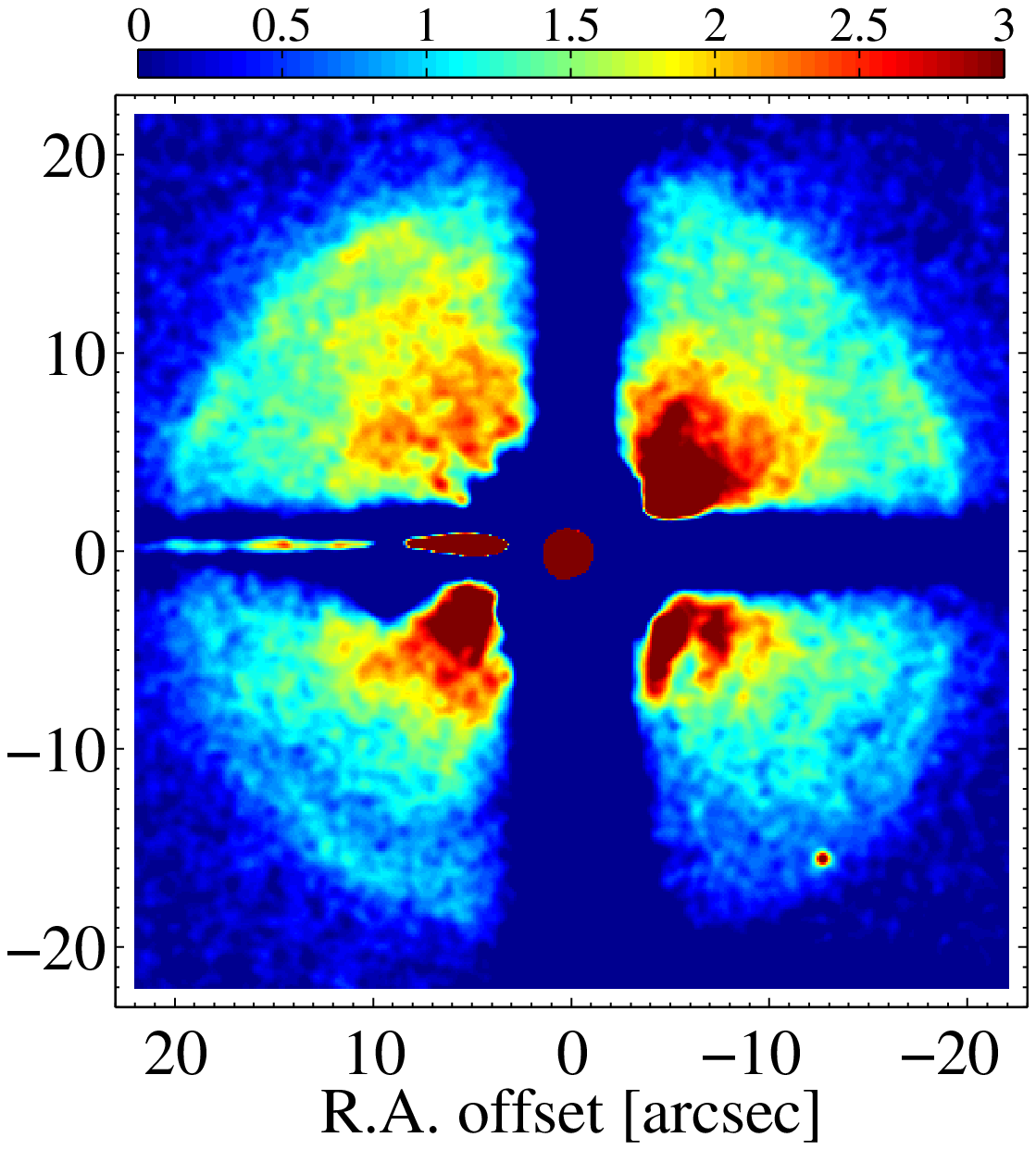}
\includegraphics[width=4.425cm]{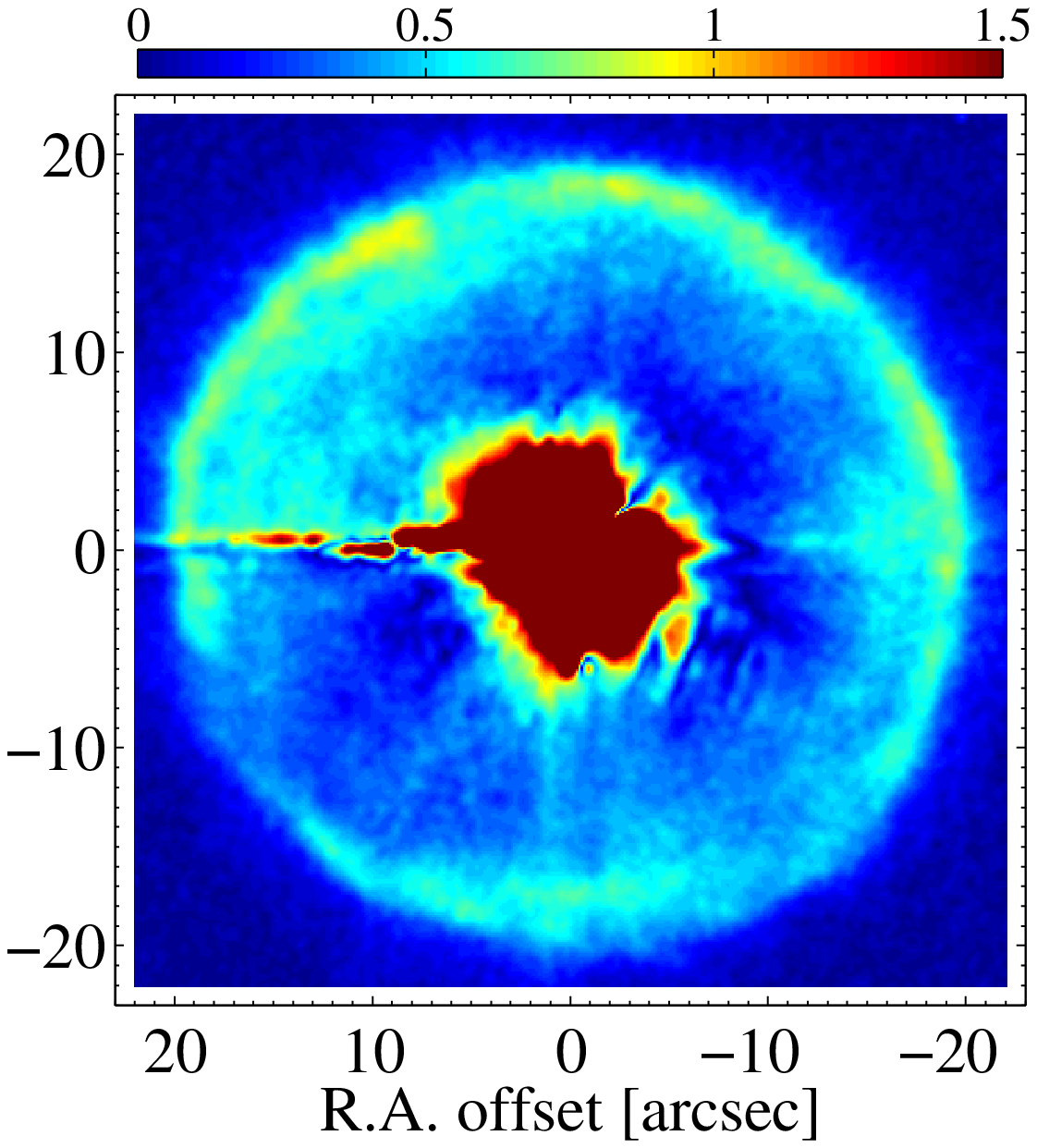}
\includegraphics[width=4.425cm]{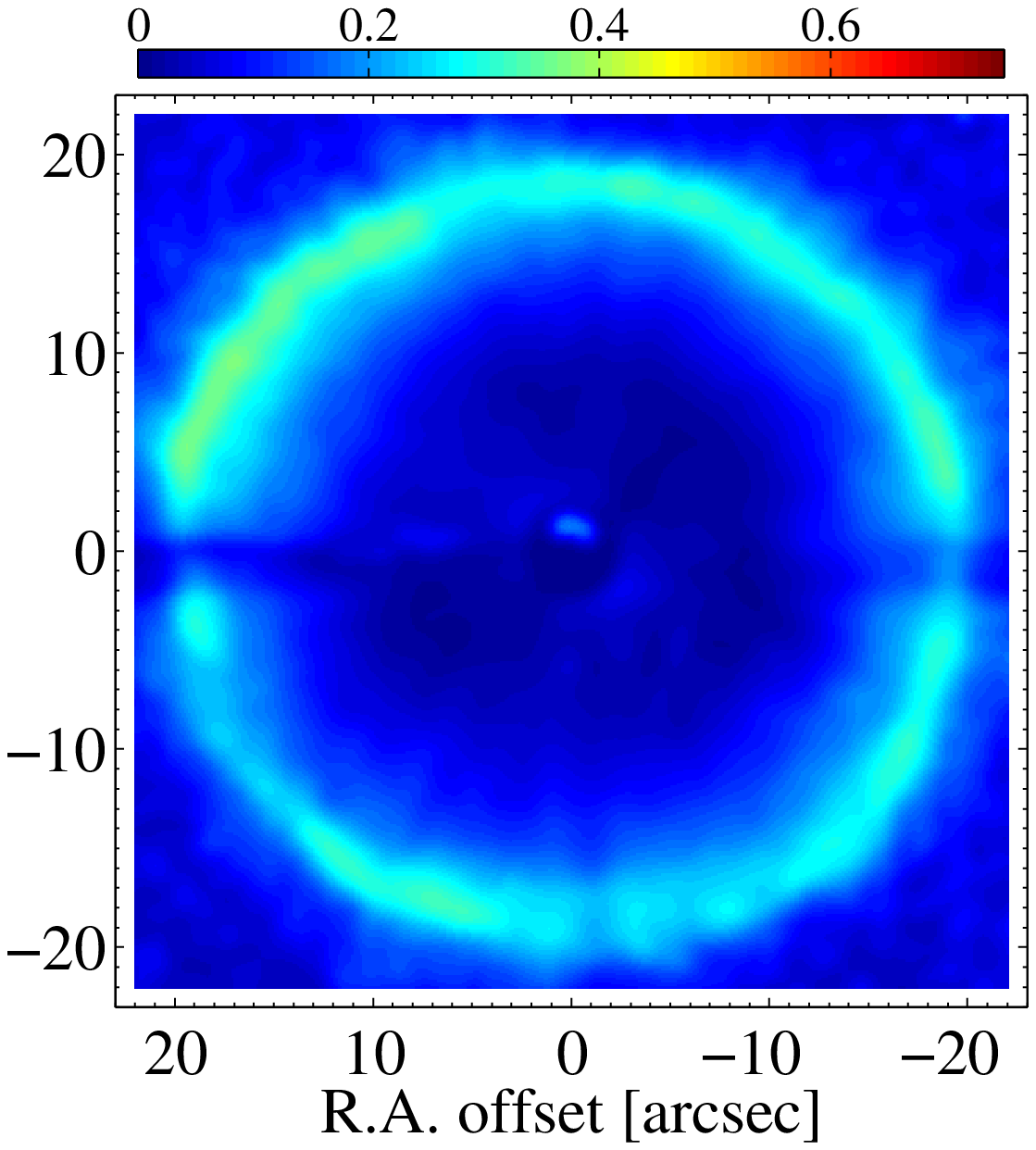}\\
\vspace{0.5cm}
\includegraphics[width=4.34cm]{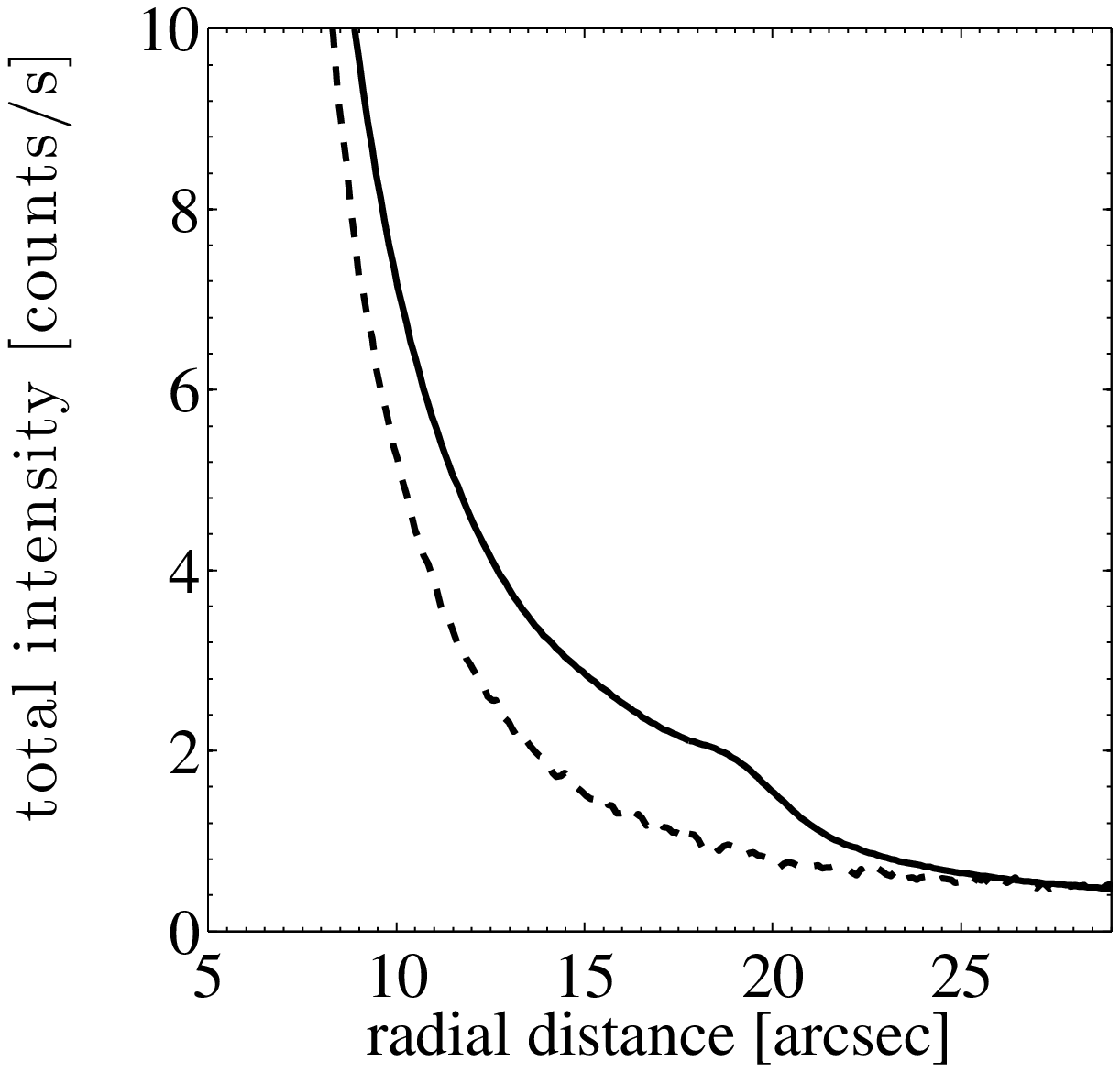}
\includegraphics[width=4.5cm]{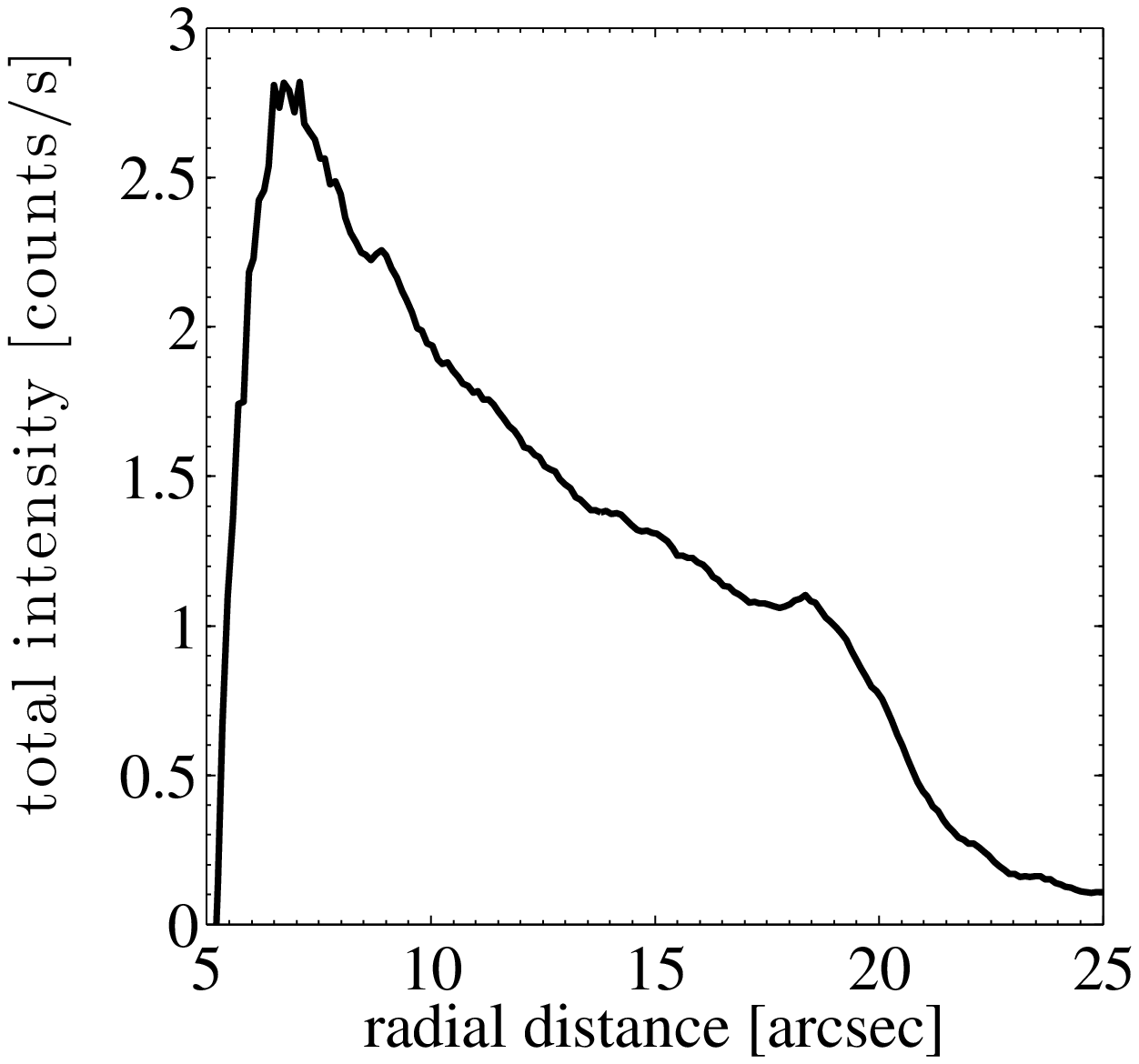}
\includegraphics[width=4.5cm]{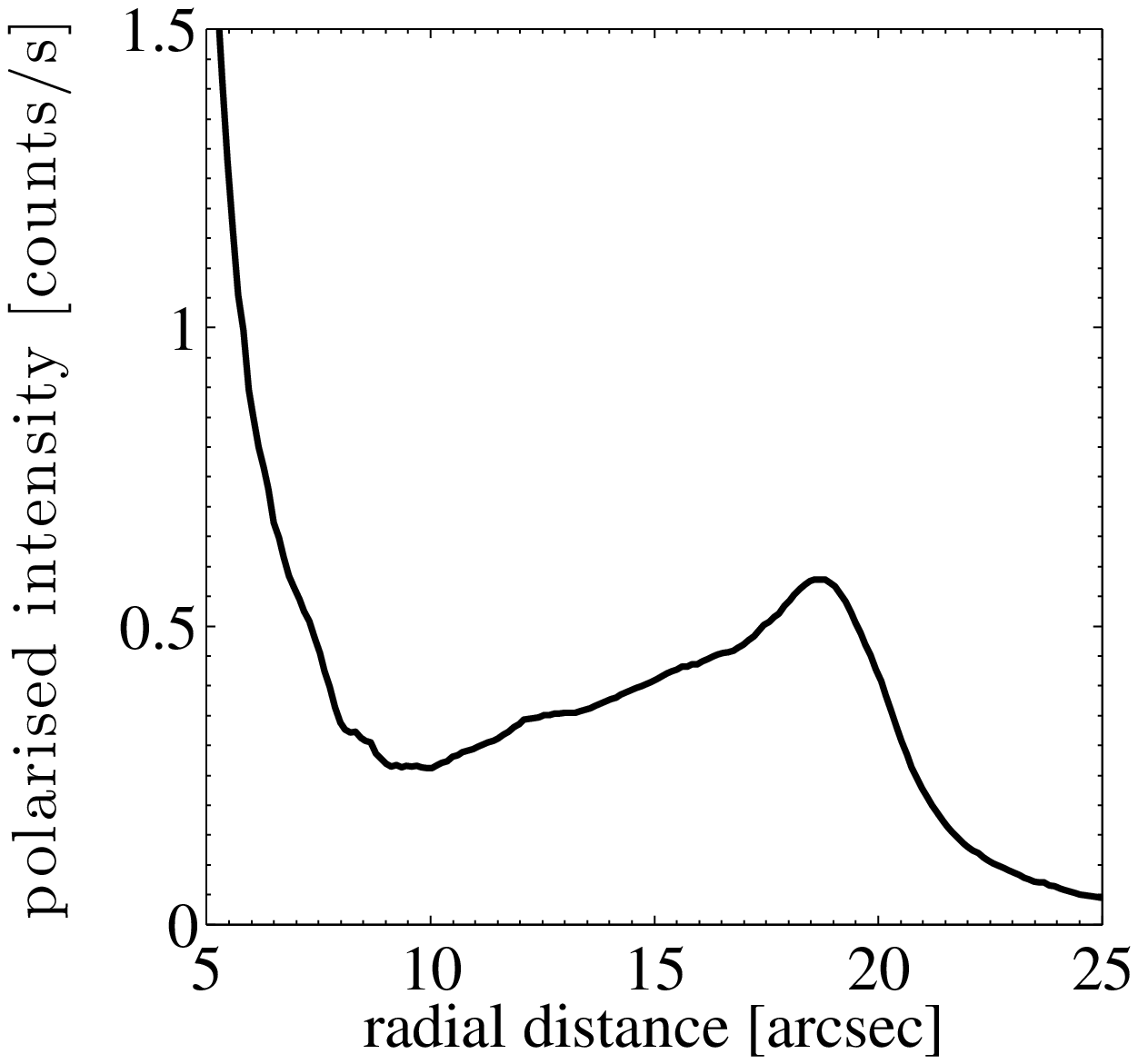}
\includegraphics[width=4.5cm]{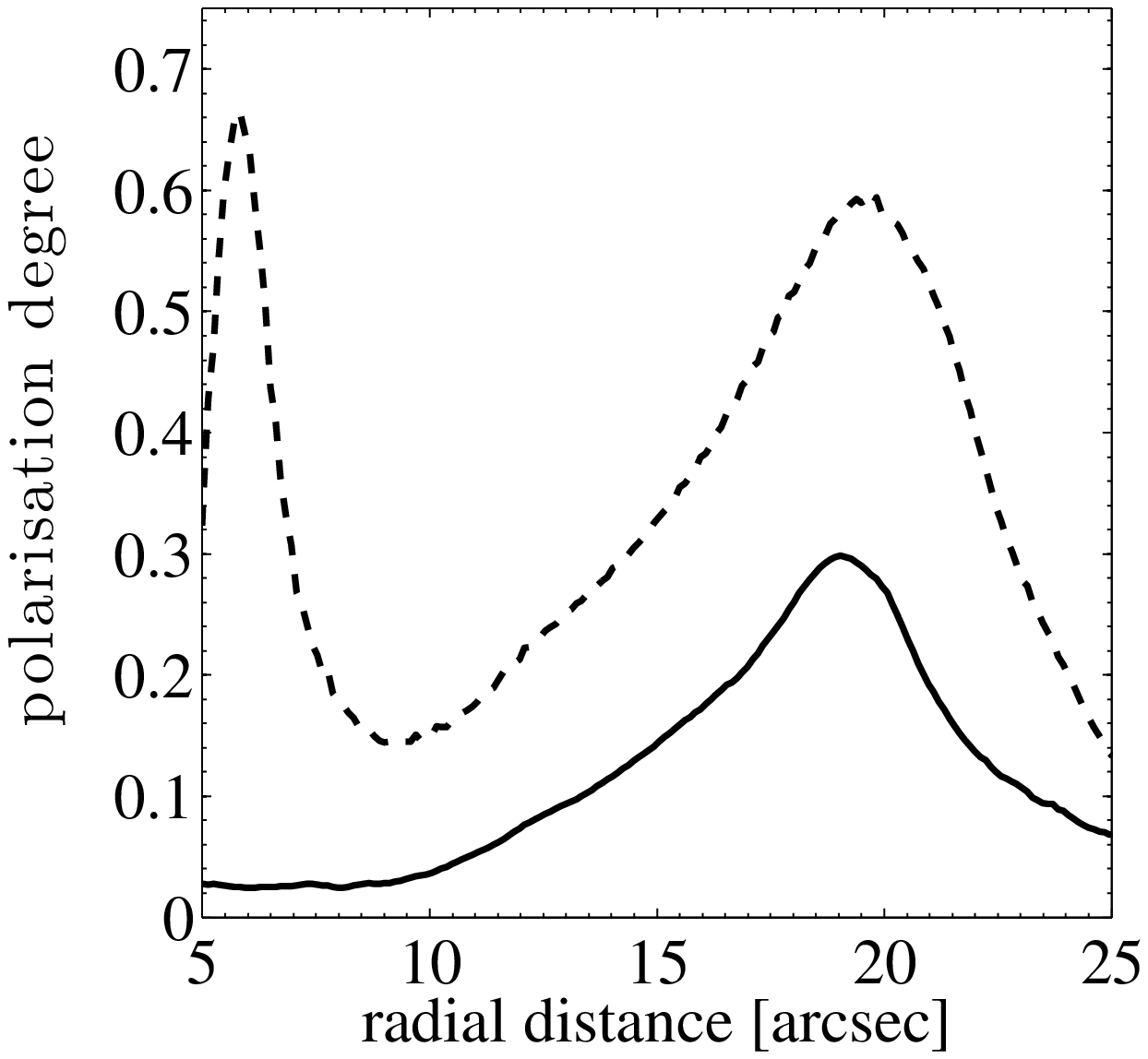}
\caption{Observations of R~Scl in the R-band. \emph{Top left to right}: total intensity, total intensity with template star subtracted, polarised intensity, and the lower limit of the polarisation degree derived using the measured total intensity (without psf subtraction). The images are smoothed by a Gaussian kernel with a FWHM of two pixels. The bottom row shows the corresponding radial profiles. The total intensity radial profile shows R~Scl (solid line) and the scaled template star (dashed line). The radial profile of the polarisation degree  shows the polarisation degree using the measured total intensity (solid line) and the template-subtracted total intensity (dashed line).}
\label{f:rsclRbandOct19}
\end{figure*}

\begin{figure*}
\centering
\includegraphics[width=4.8cm]{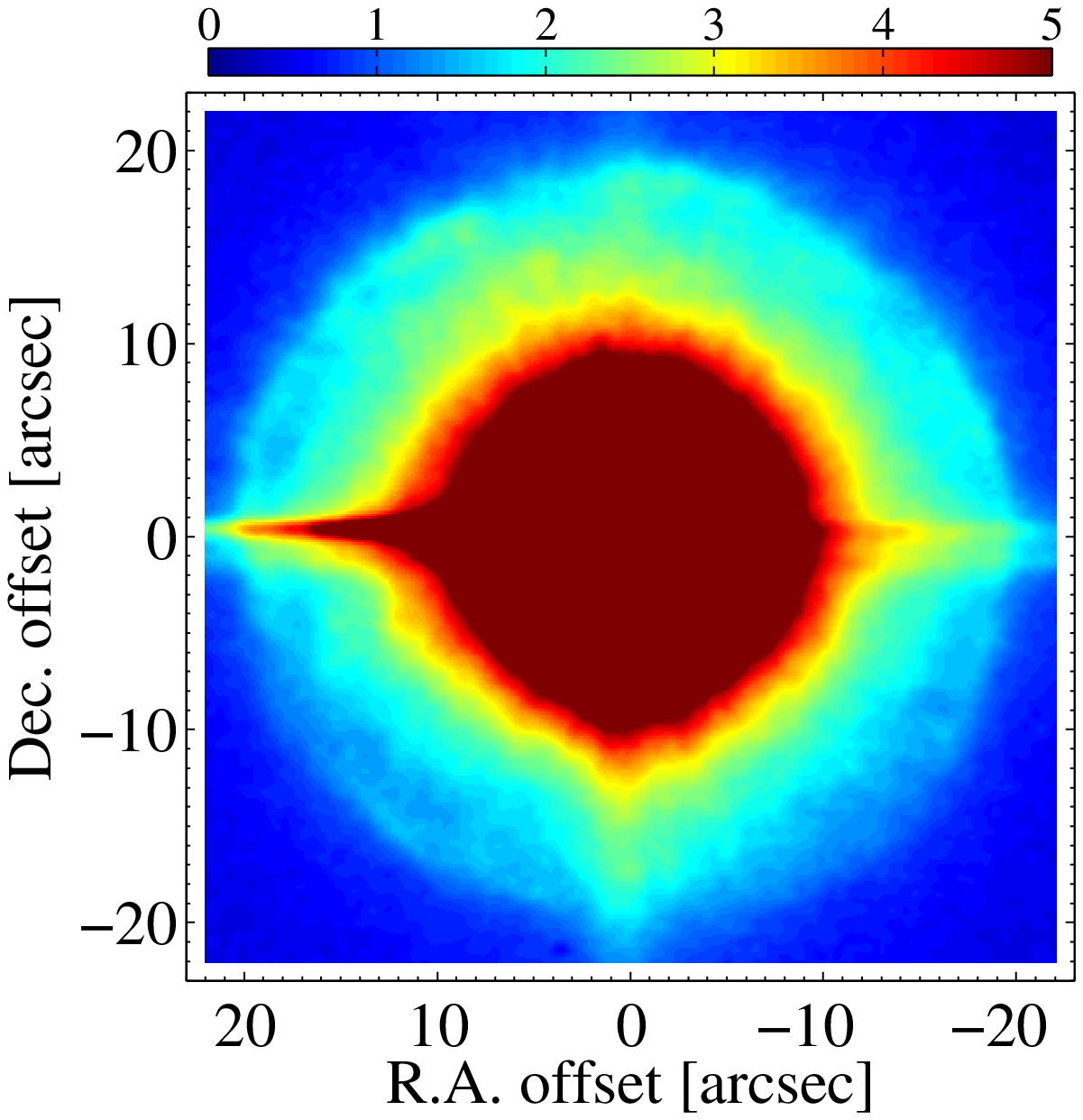}
\includegraphics[width=4.425cm]{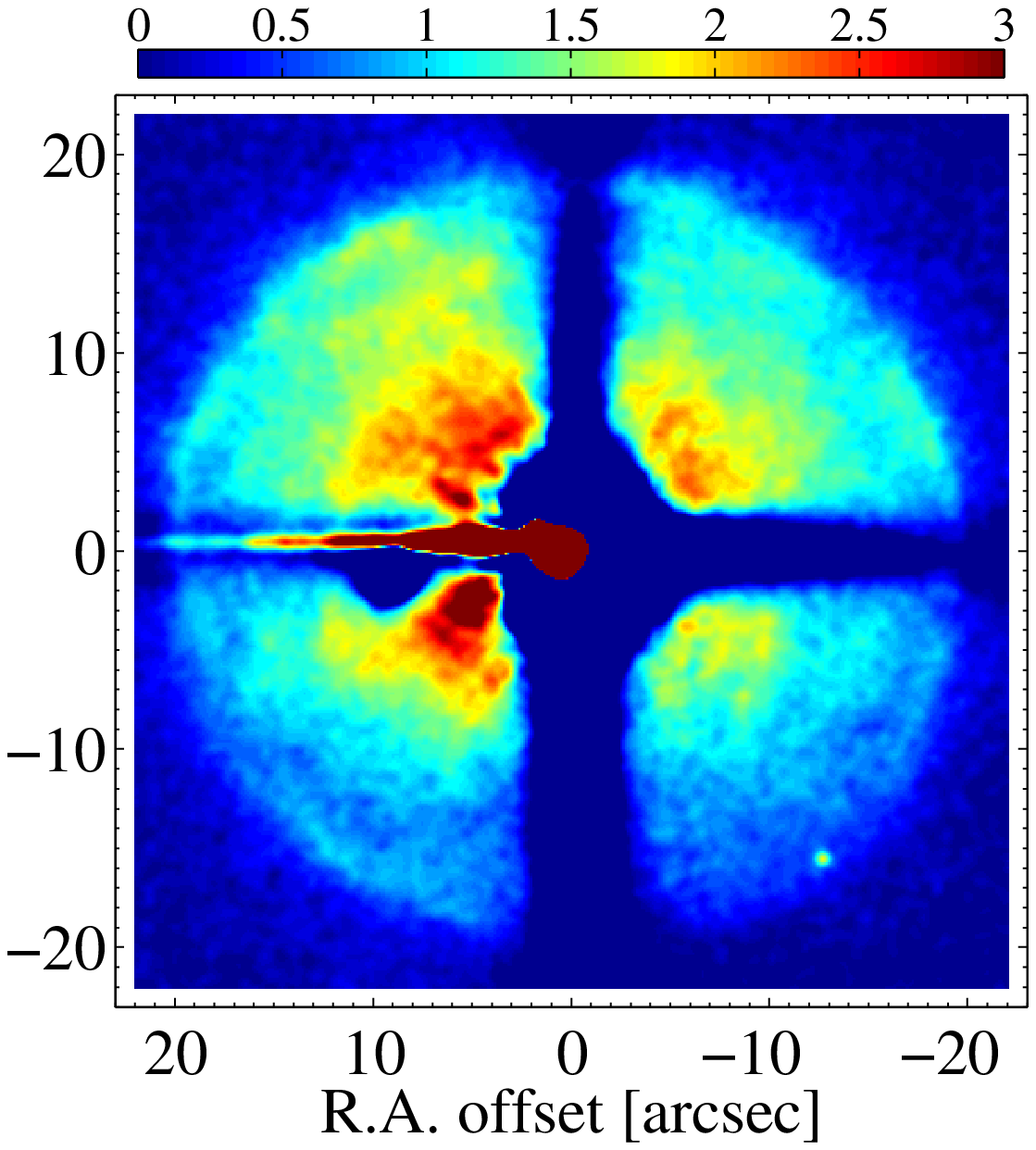}
\includegraphics[width=4.425cm]{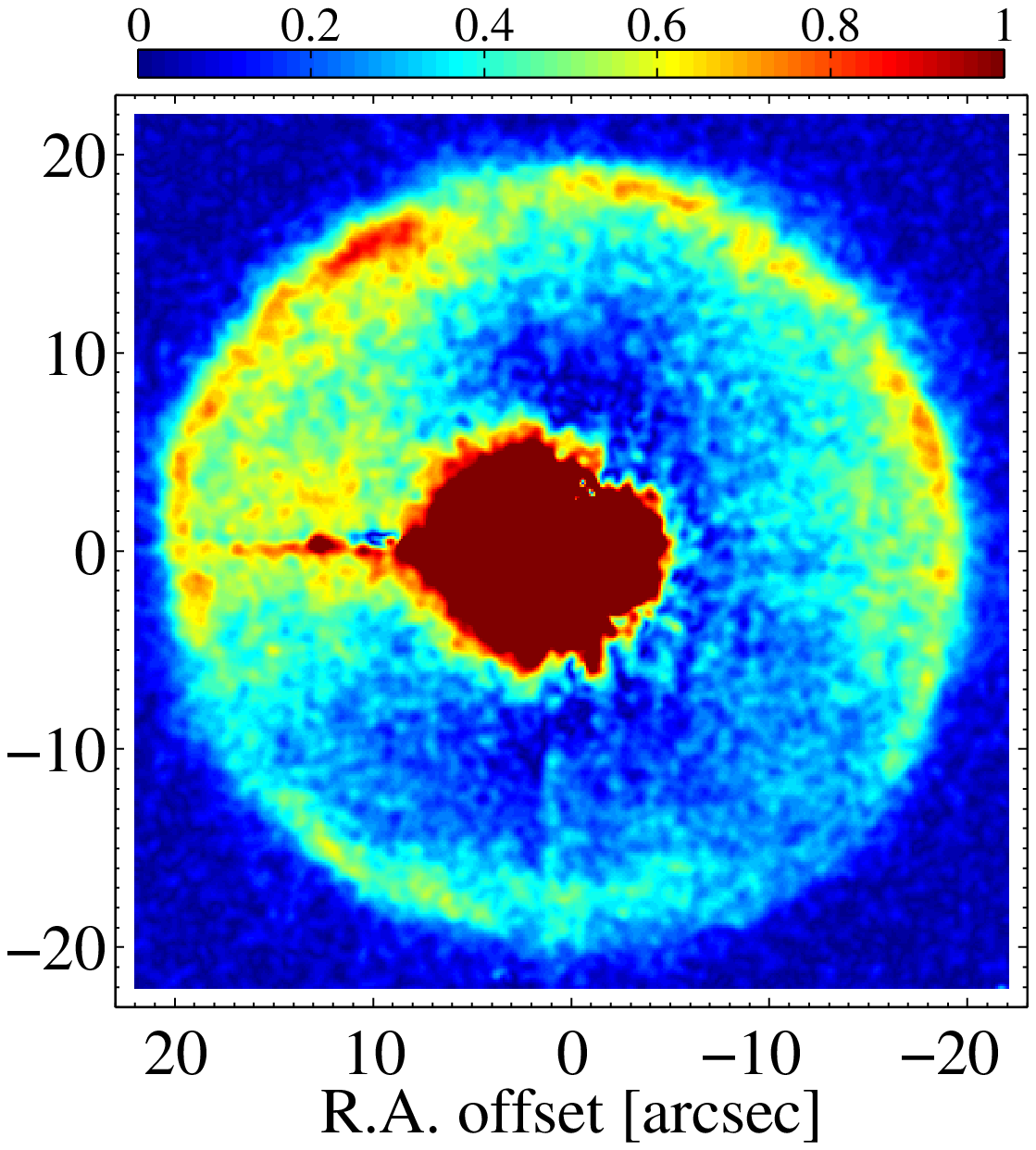}
\includegraphics[width=4.425cm]{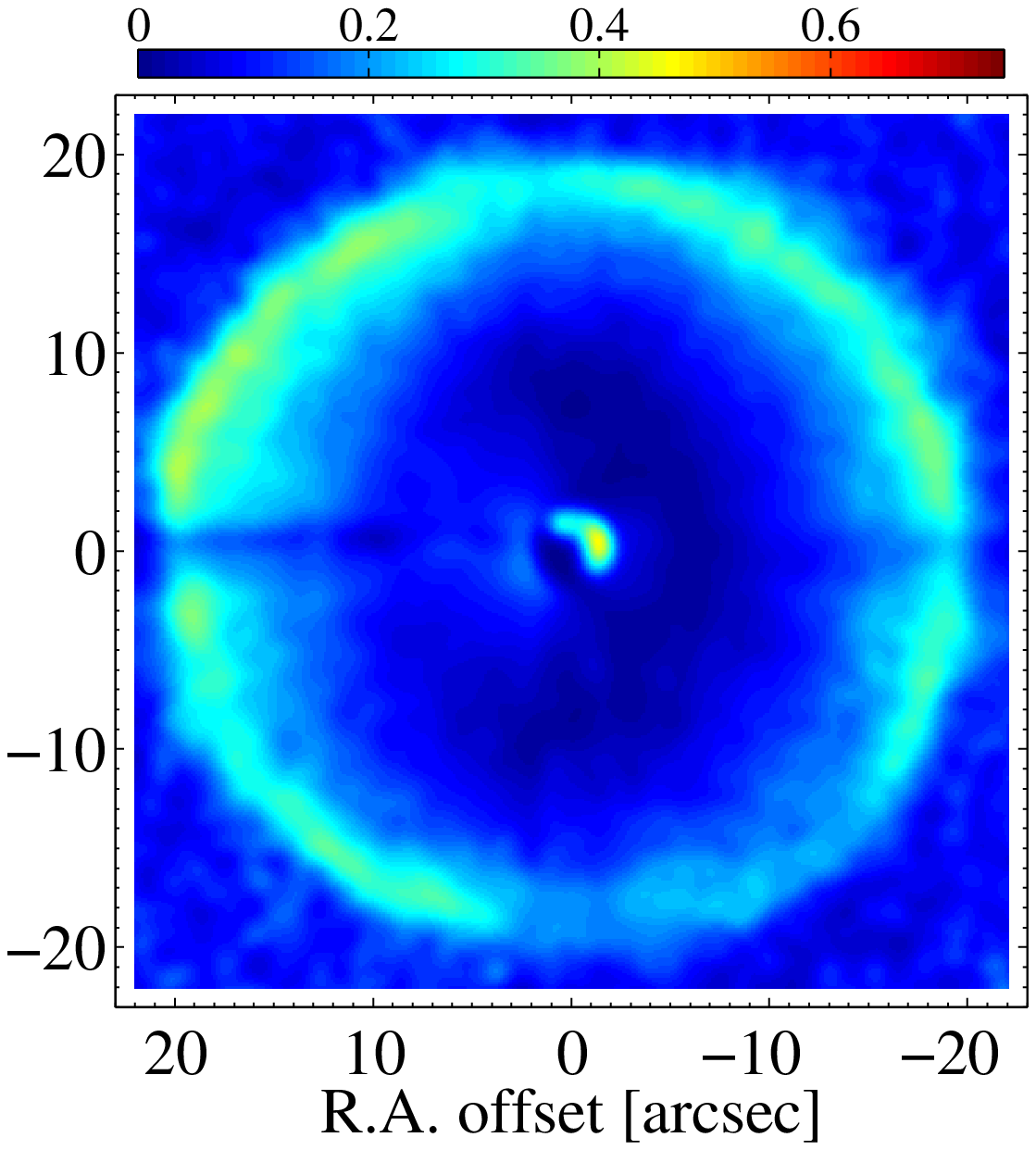}\\
\vspace{0.5cm}
\includegraphics[width=4.34cm]{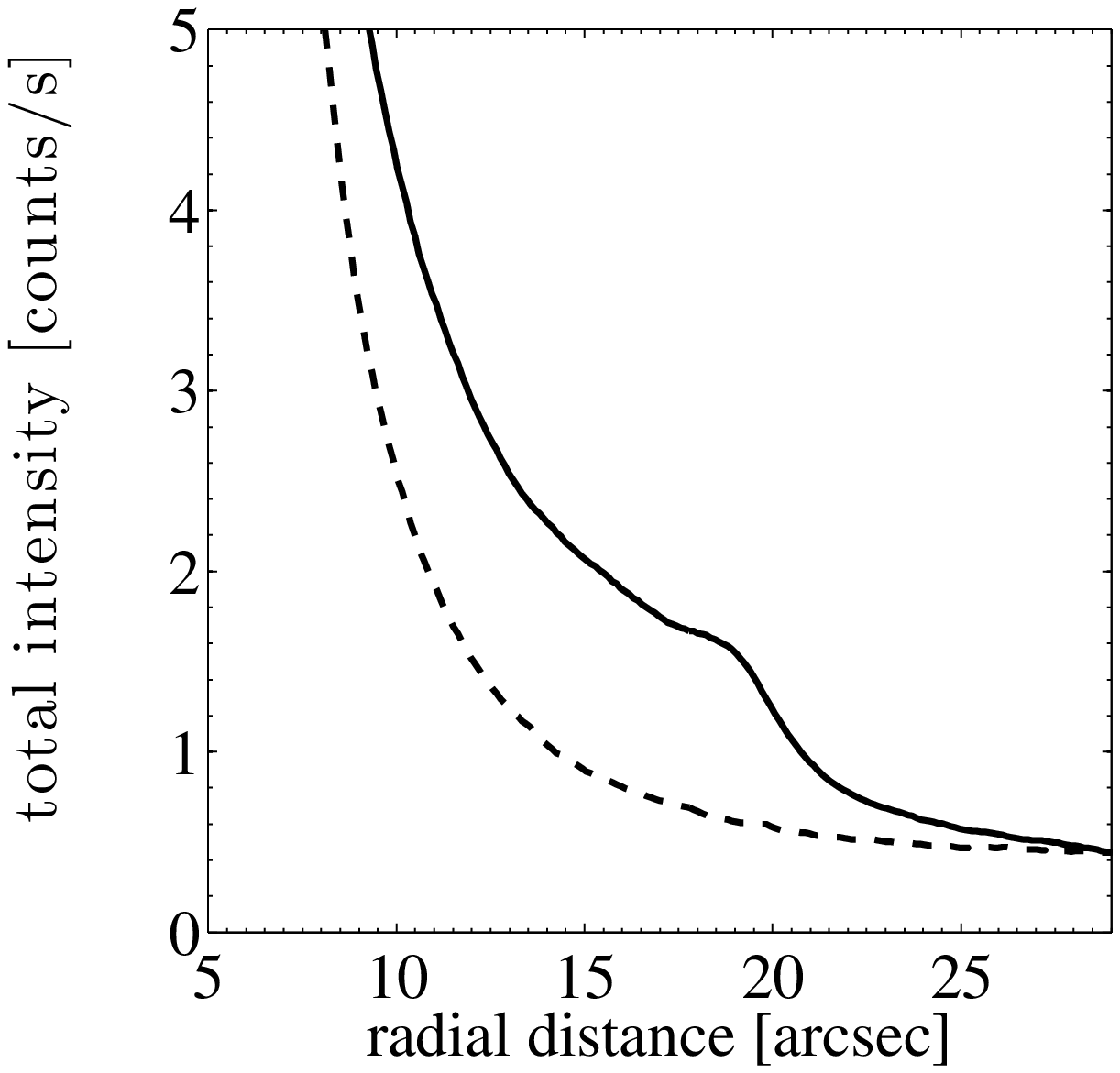}
\includegraphics[width=4.5cm]{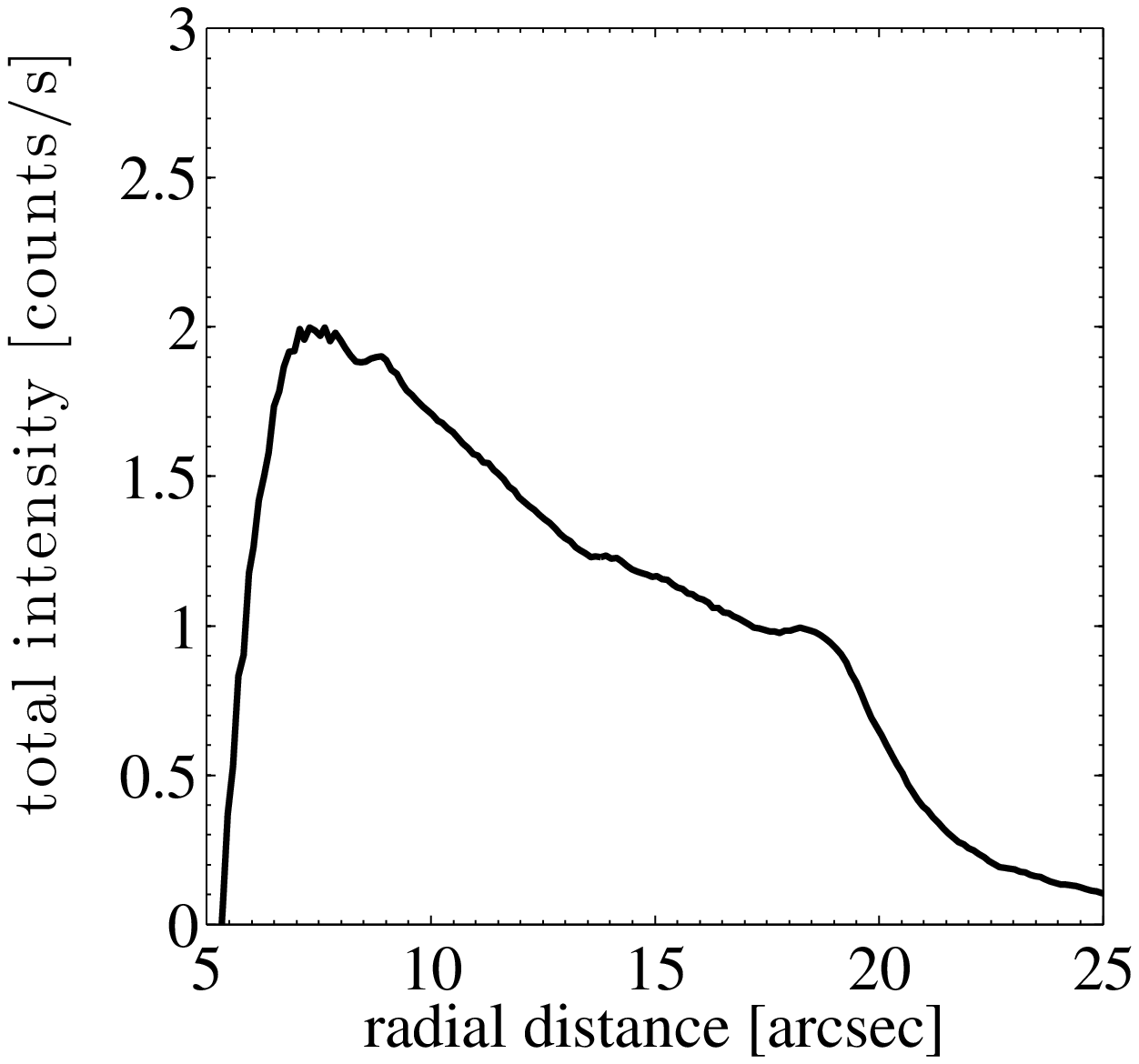}
\includegraphics[width=4.5cm]{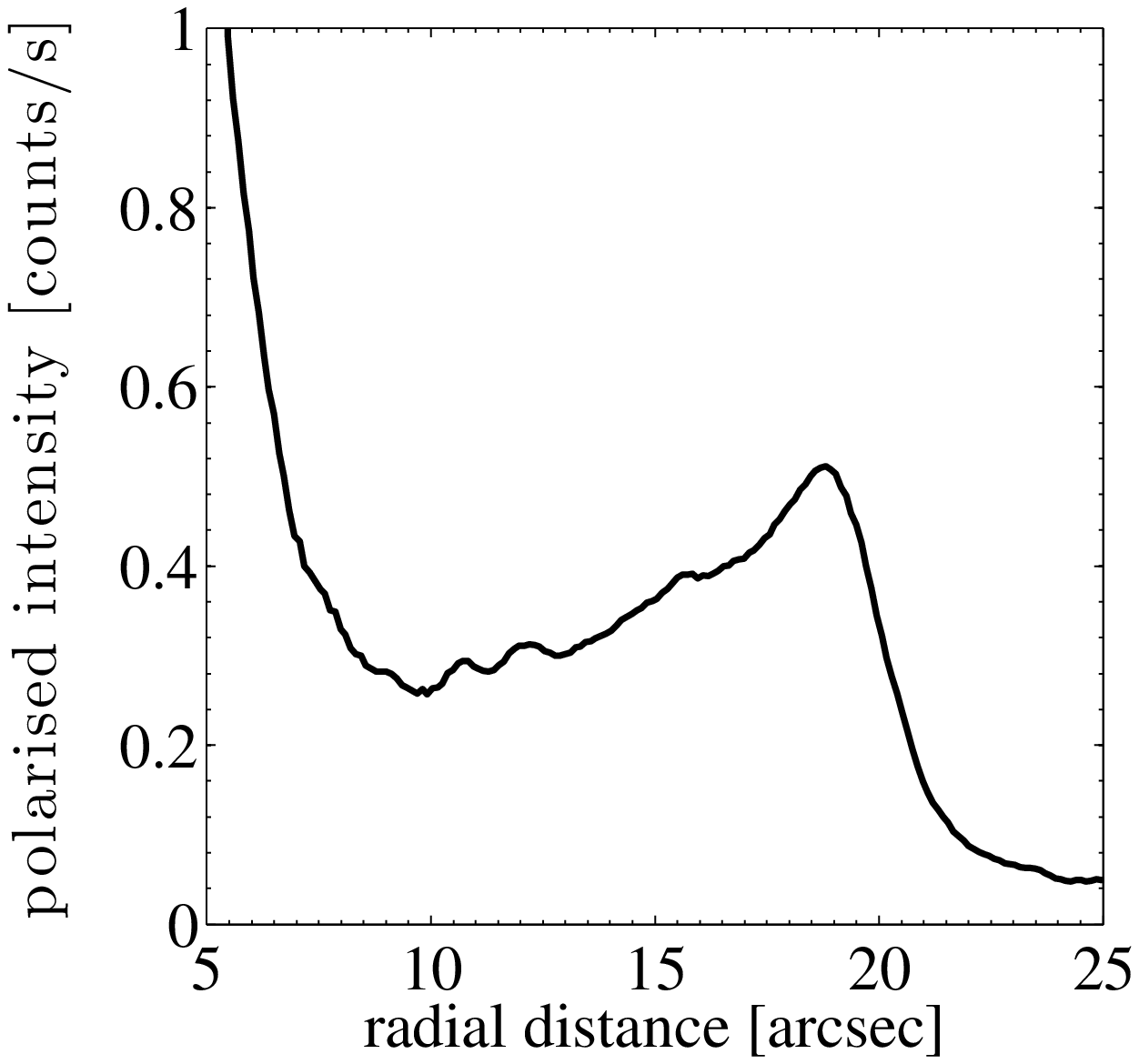}
\includegraphics[width=4.5cm]{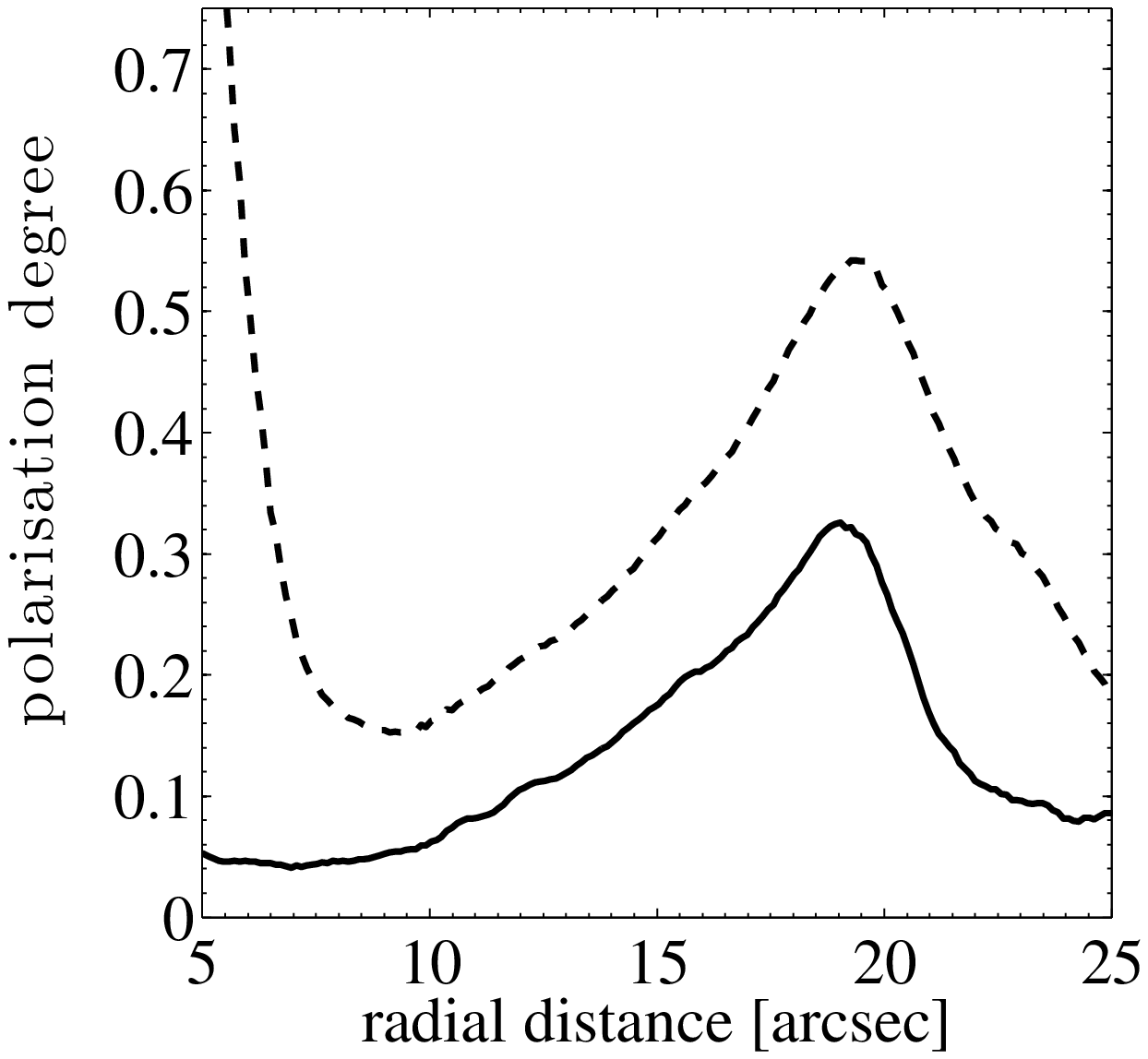}
\caption{Same as Fig.~\ref{f:rsclRbandOct19} but in the V-band.}
\label{f:rsclVbandOct19}
\end{figure*}

\begin{figure*}
\centering
\includegraphics[width=4.8cm]{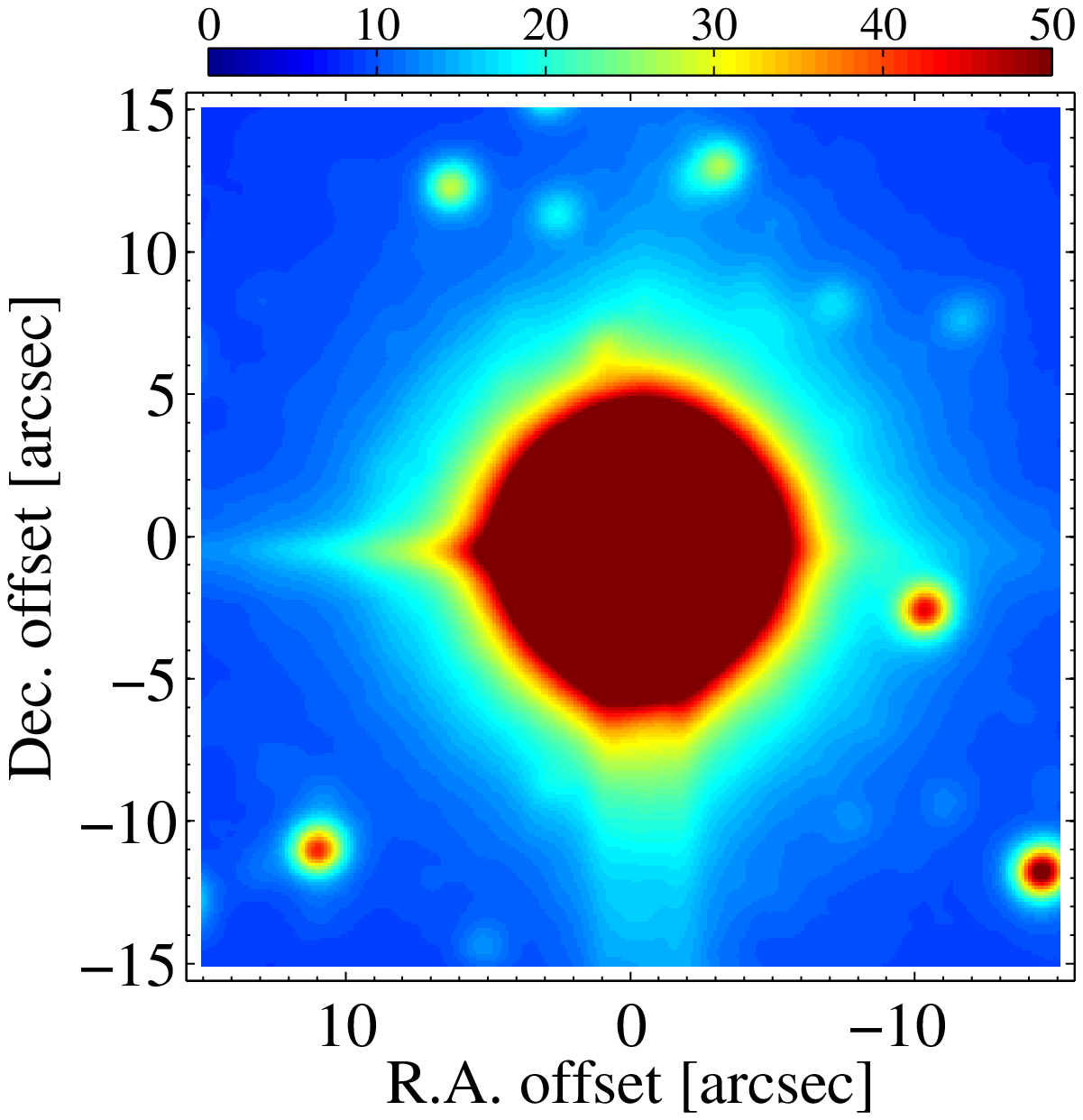}
\includegraphics[width=4.425cm]{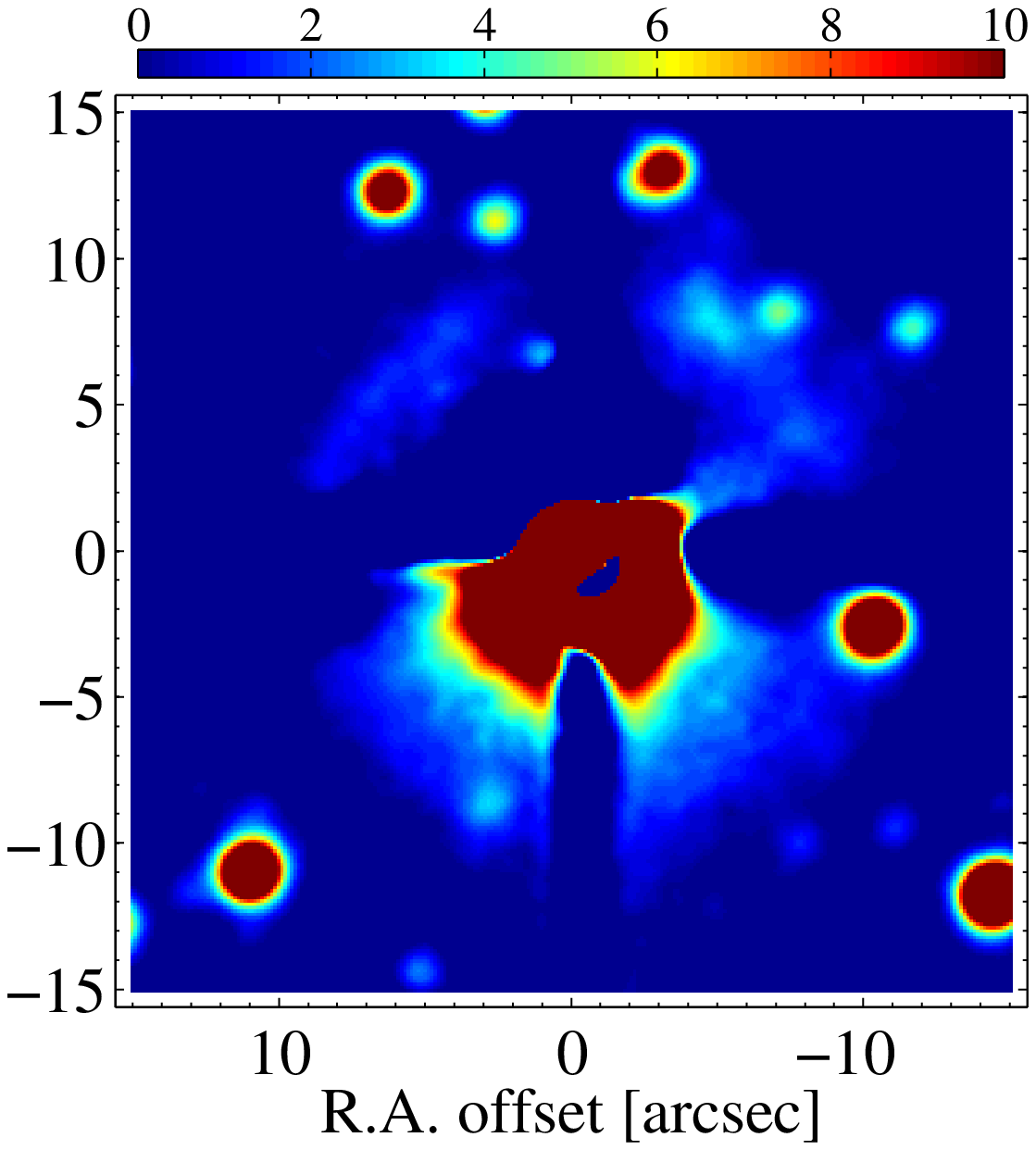}
\includegraphics[width=4.425cm]{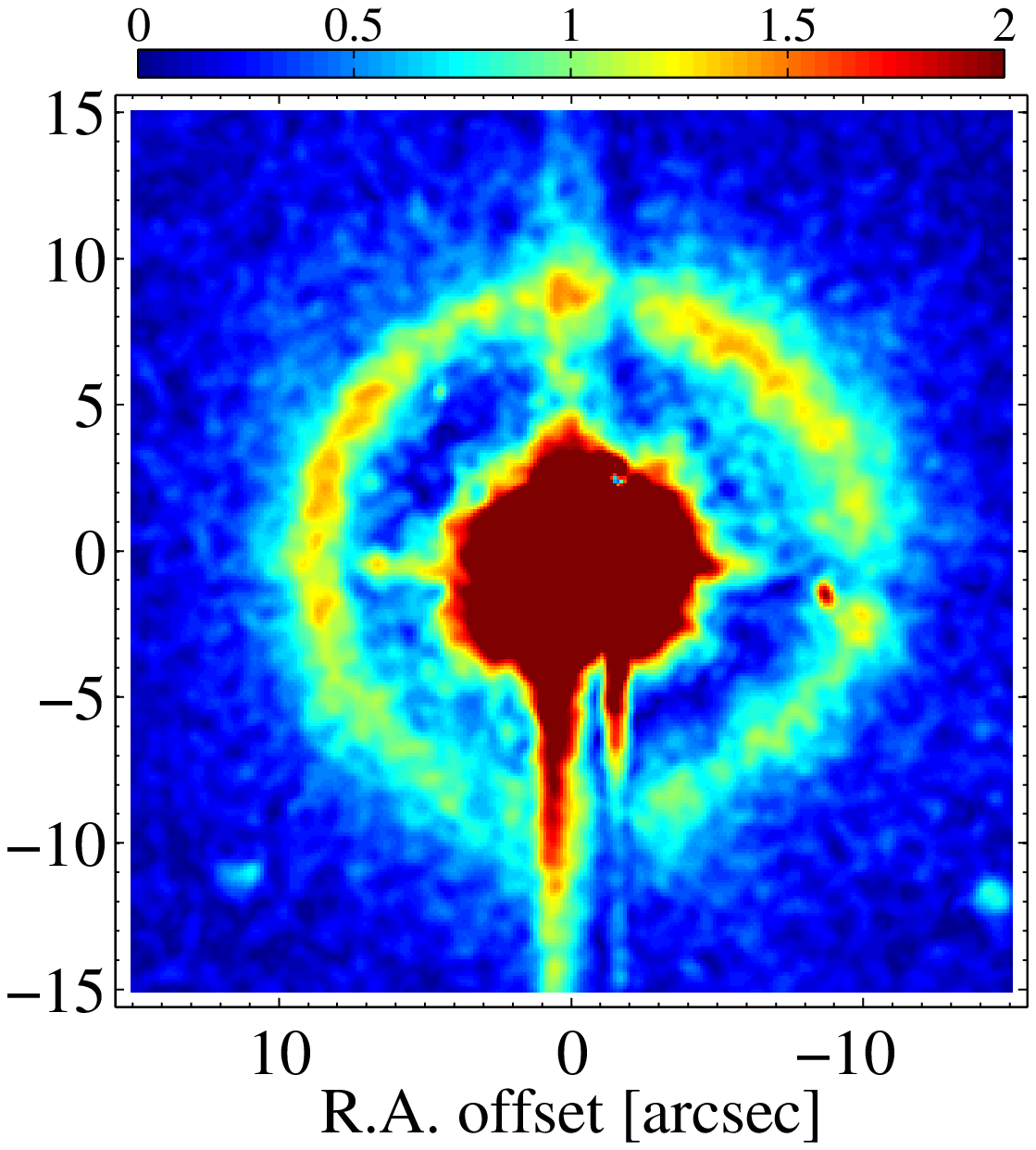}
\includegraphics[width=4.425cm]{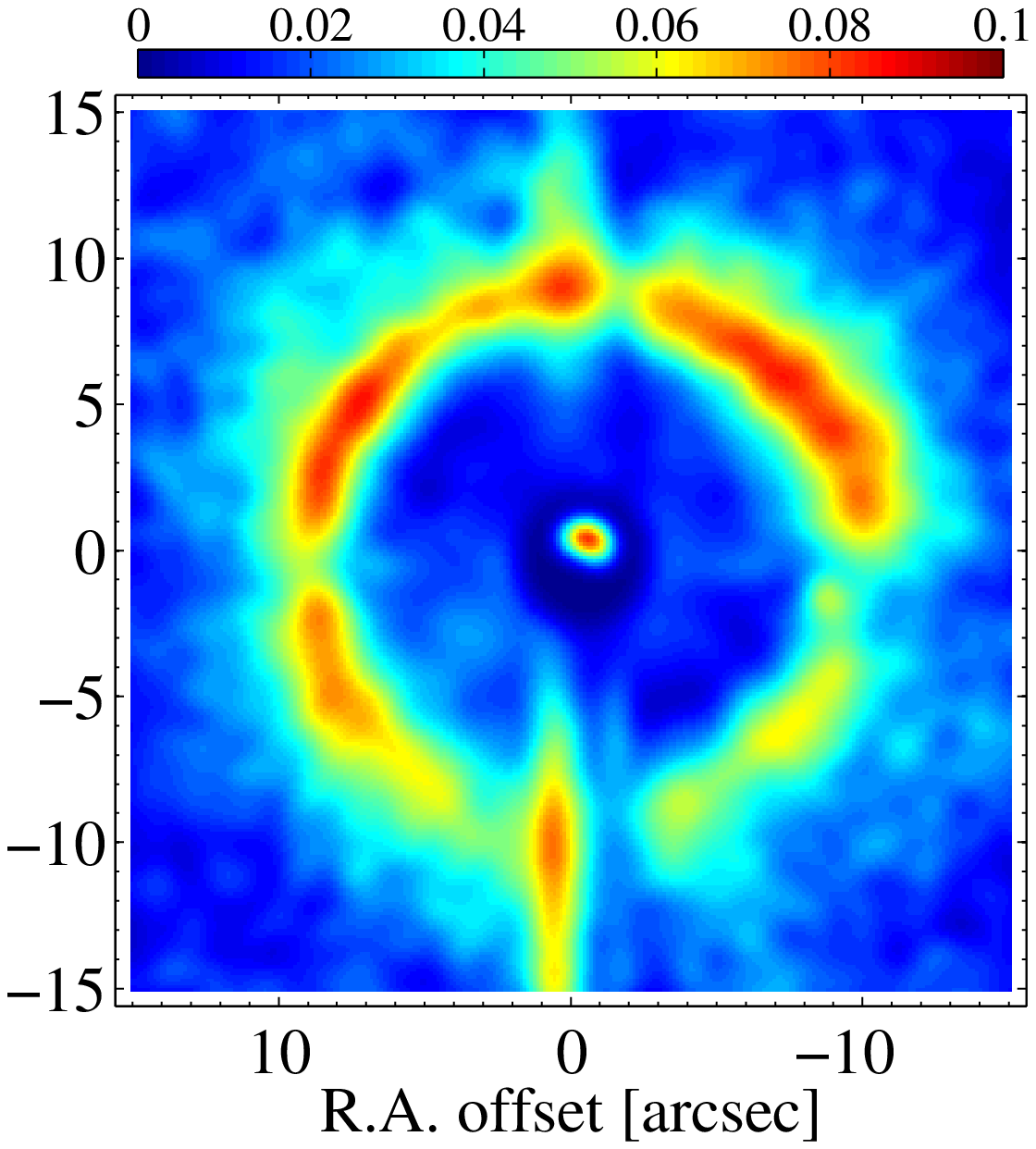}\\
\vspace{0.5cm}
\includegraphics[width=4.45cm]{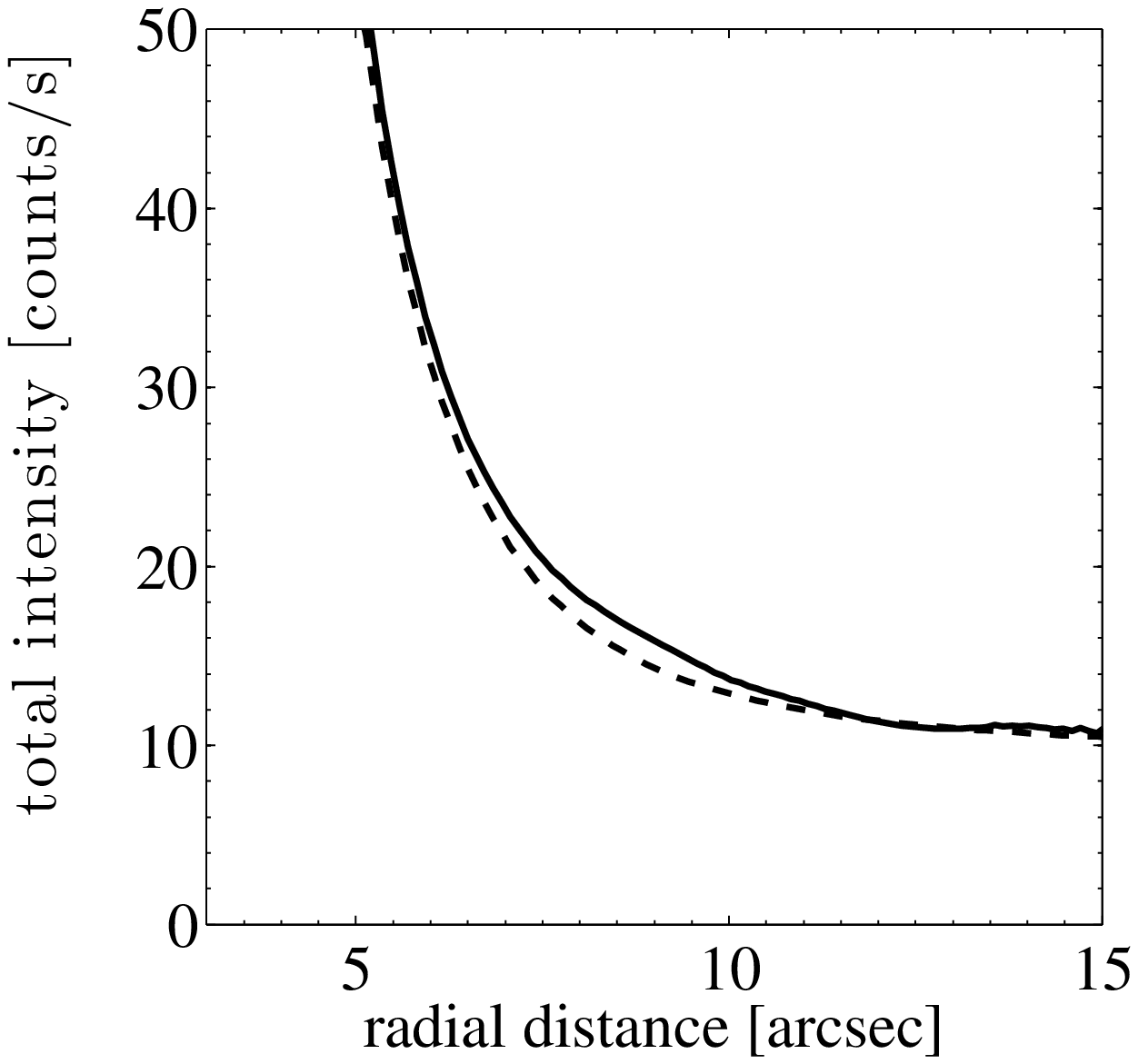}
\includegraphics[width=4.5cm]{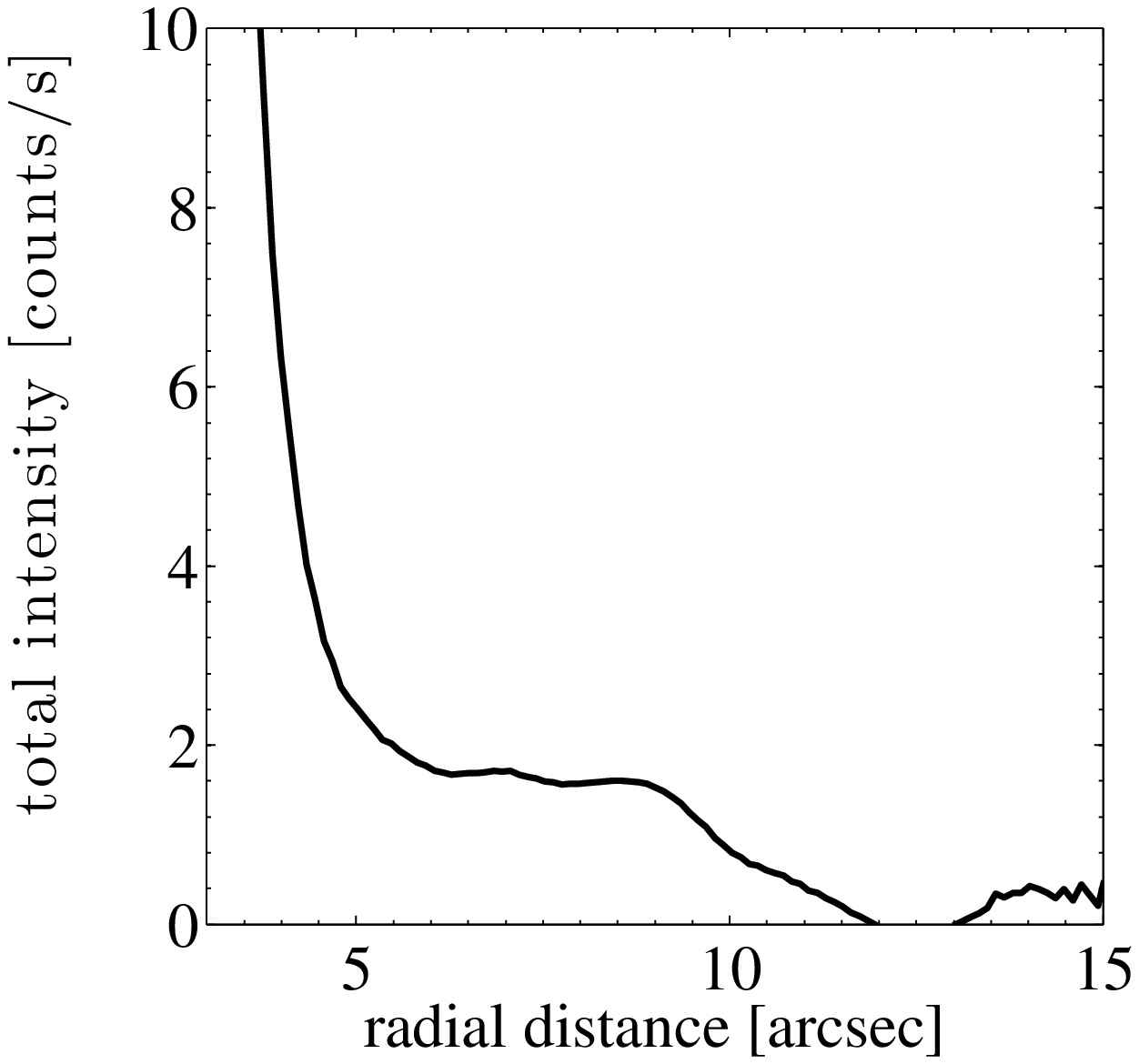}
\includegraphics[width=4.5cm]{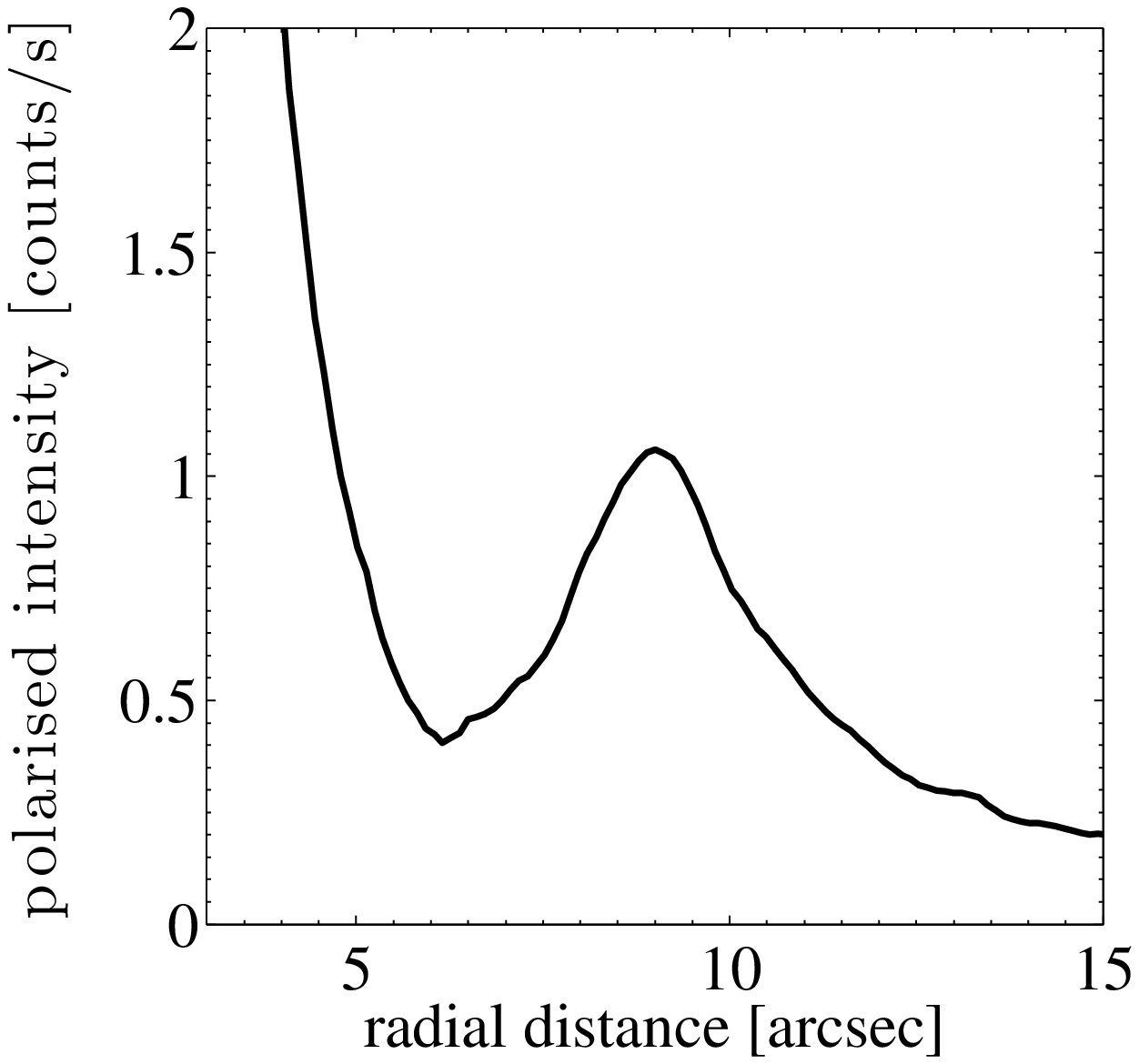}
\includegraphics[width=4.5cm]{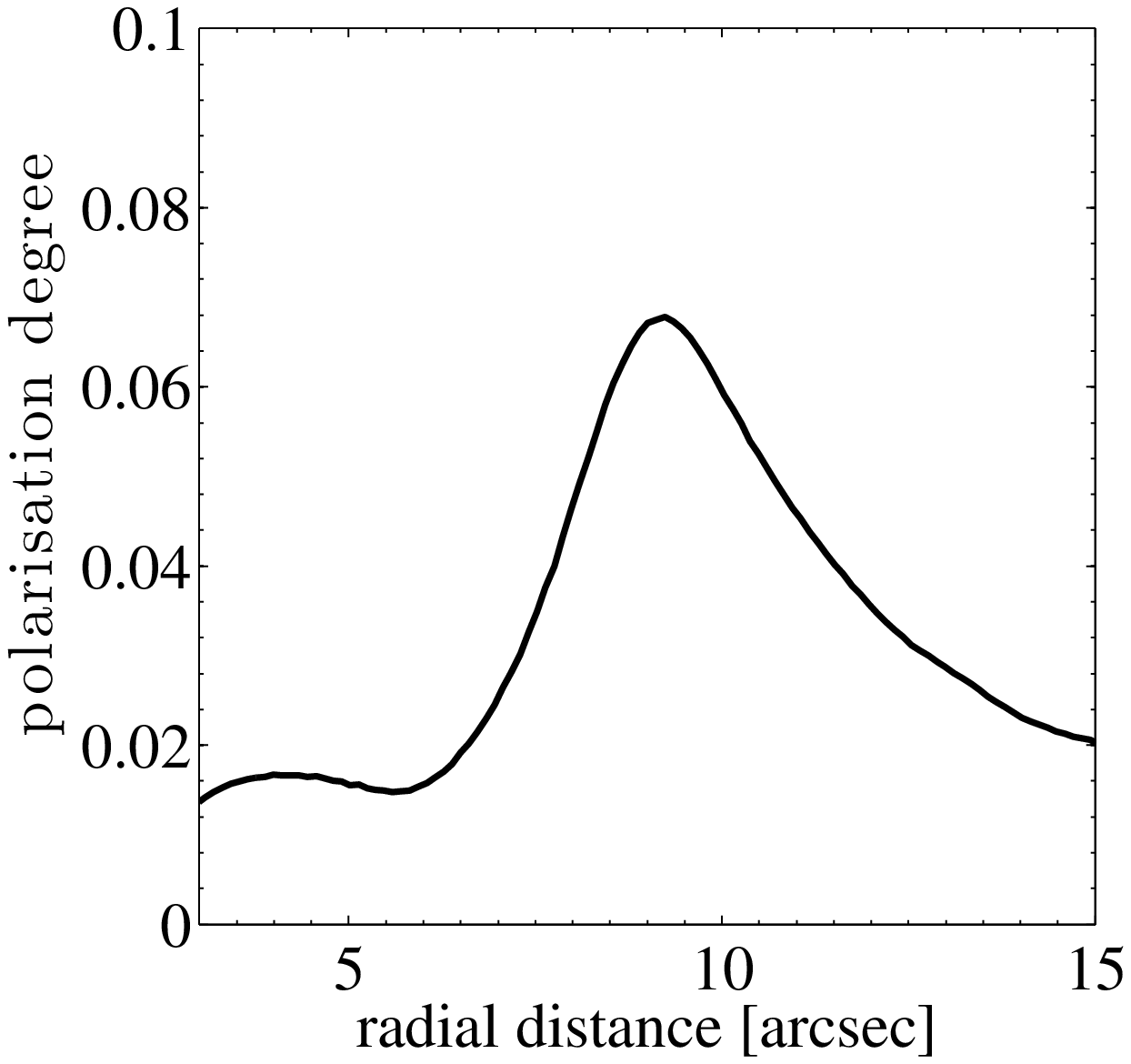}
\caption{Observations of V644~Sco in the R-band. \emph{Top left to right}: total intensity, total intensity with template star subtracted, polarised intensity, and the lower limit of the polarisation degree derived using the measured total intensity (without psf subtraction). The images are smoothed by a Gaussian kernel with a FWHM of two pixels. The bottom row shows the corresponding radial profiles. The total intensity radial profile shows V644~Sco (solid line) and the scaled template star (dashed line). The radial profile of the polarisation degree shows the polarisation degree using the measured total intensity only (solid line).}
\label{f:v644RbandOct24}
\end{figure*}

\begin{figure*}
\centering
\includegraphics[width=4.8cm]{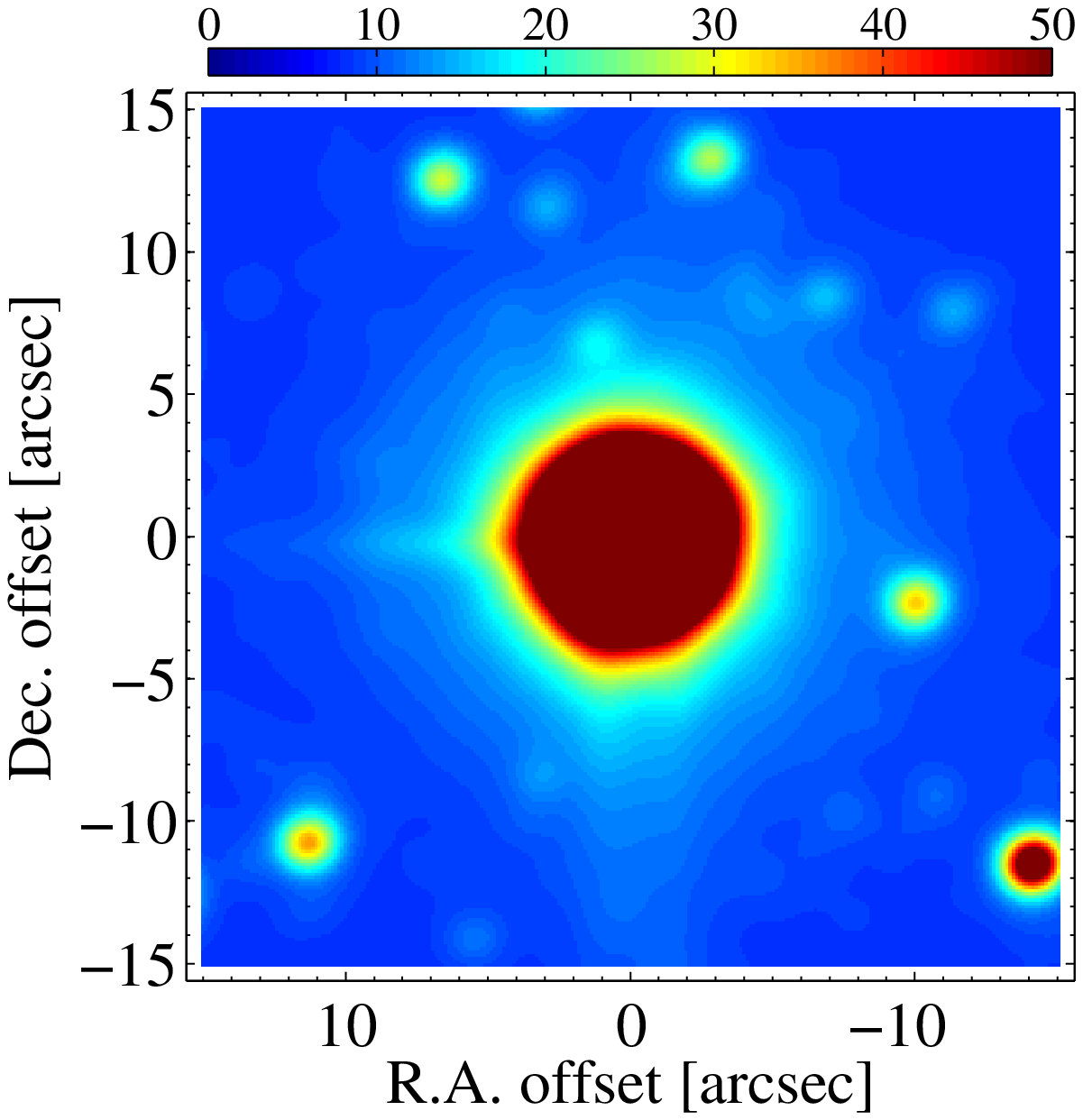}
\includegraphics[width=4.425cm]{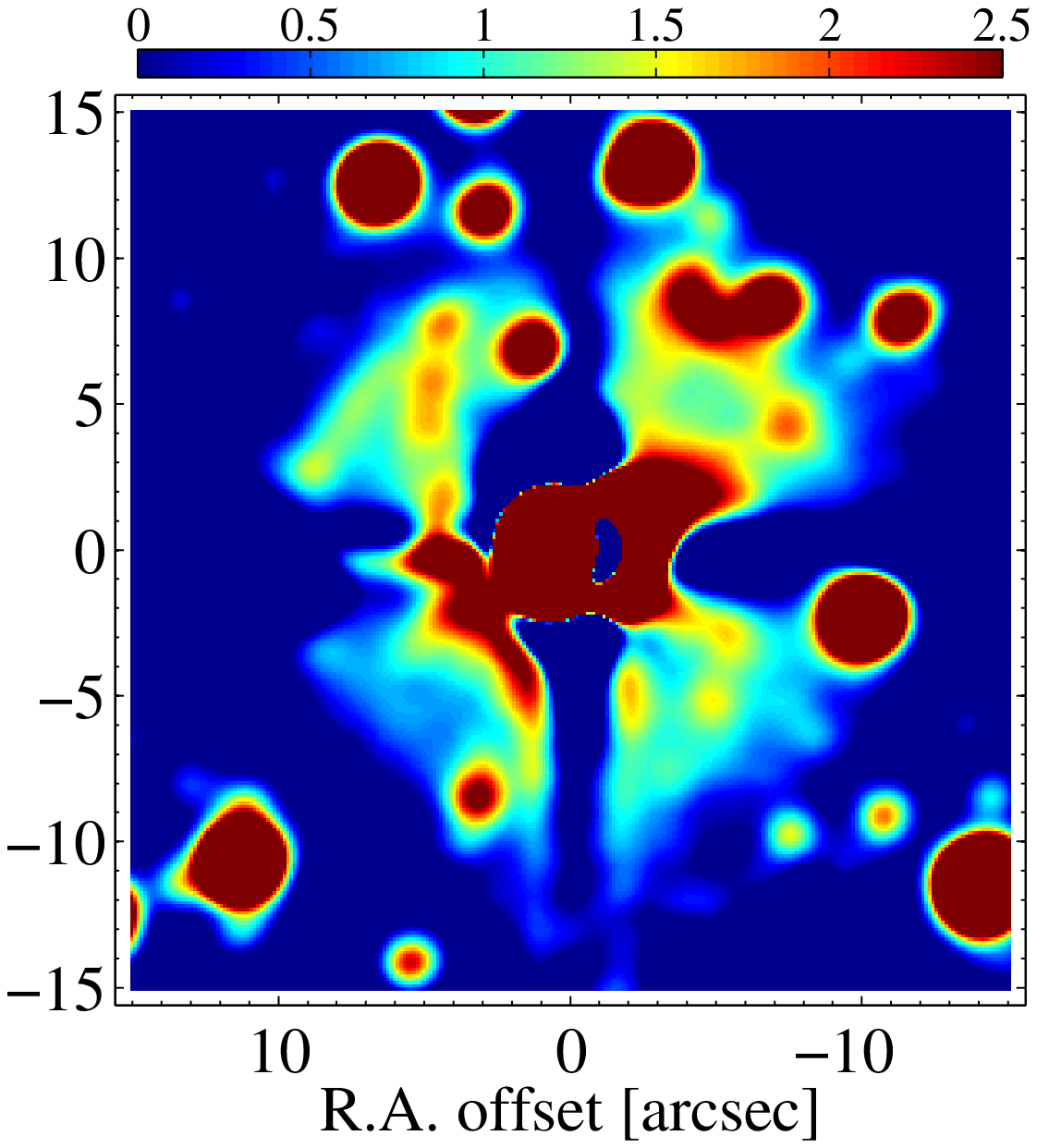}
\includegraphics[width=4.425cm]{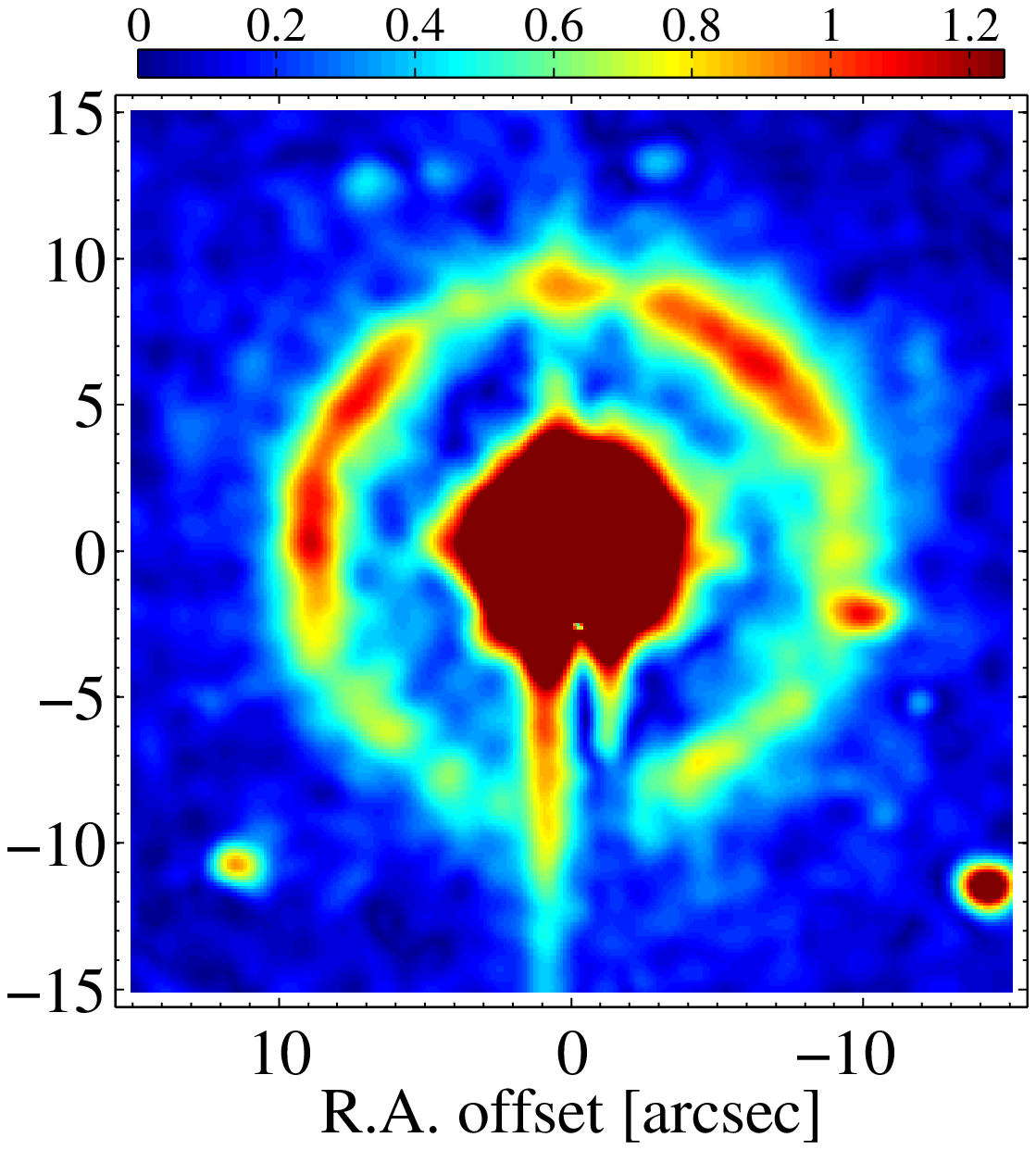}
\includegraphics[width=4.425cm]{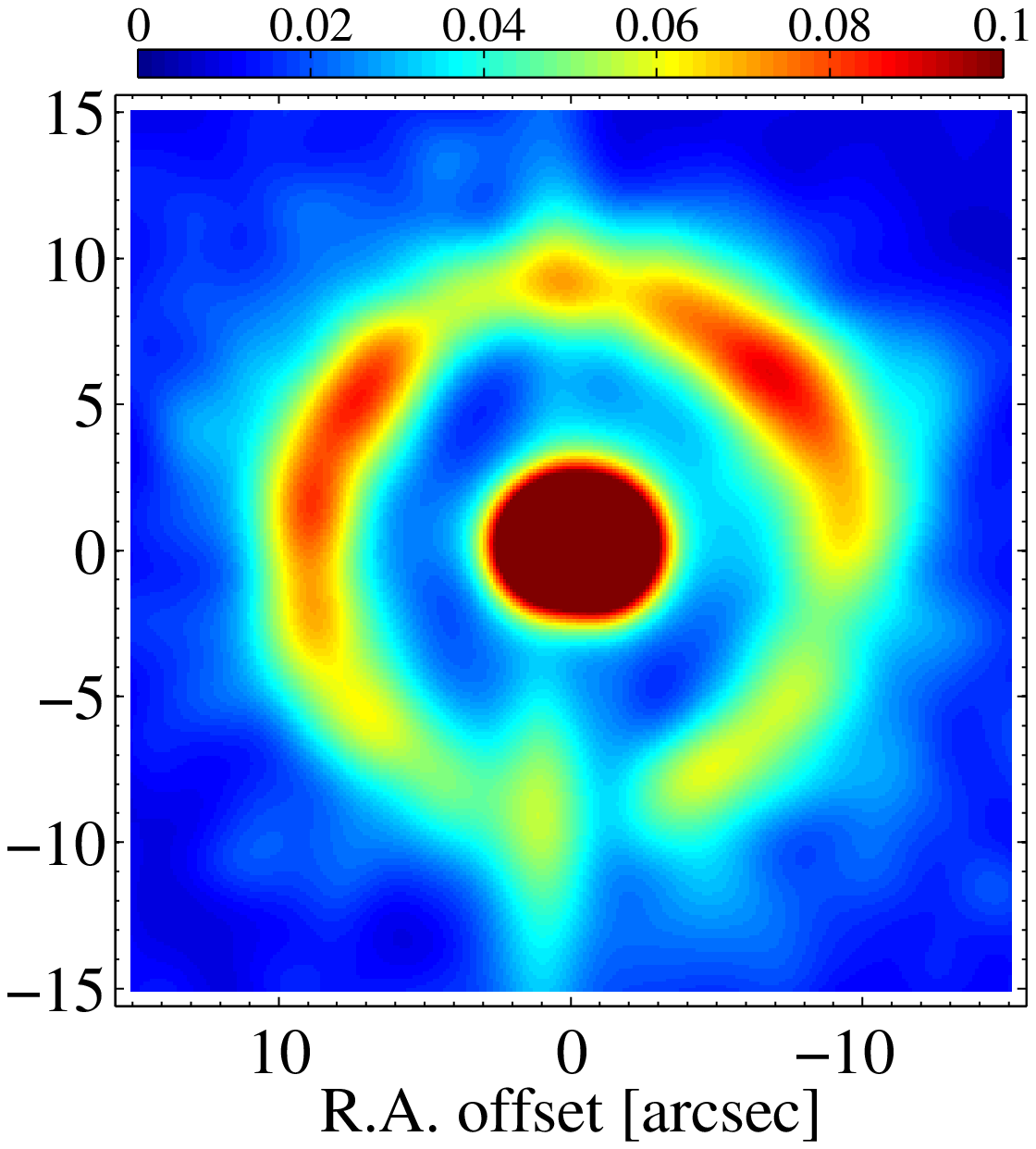}\\
\vspace{0.5cm}
\includegraphics[width=4.45cm]{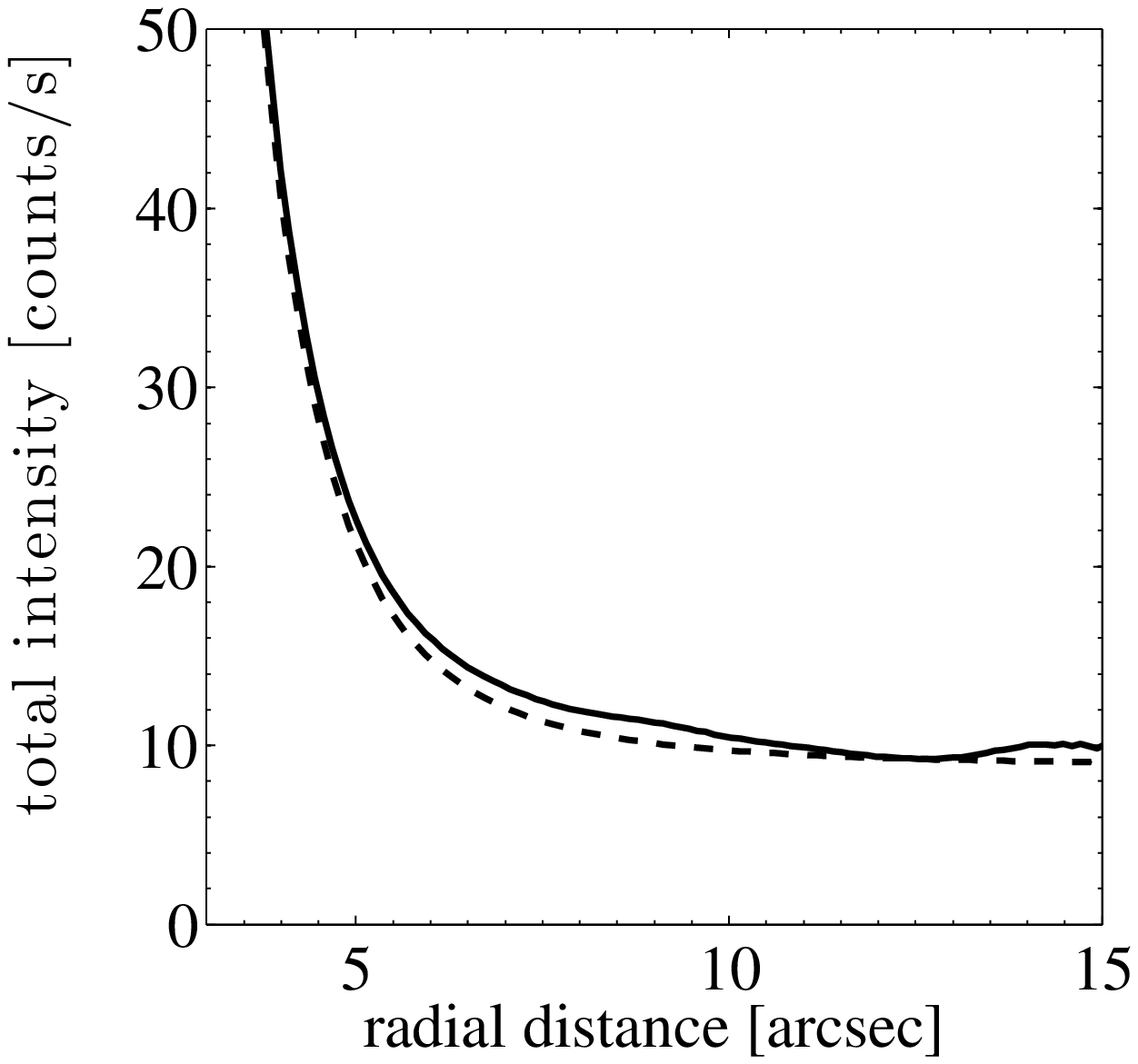}
\includegraphics[width=4.5cm]{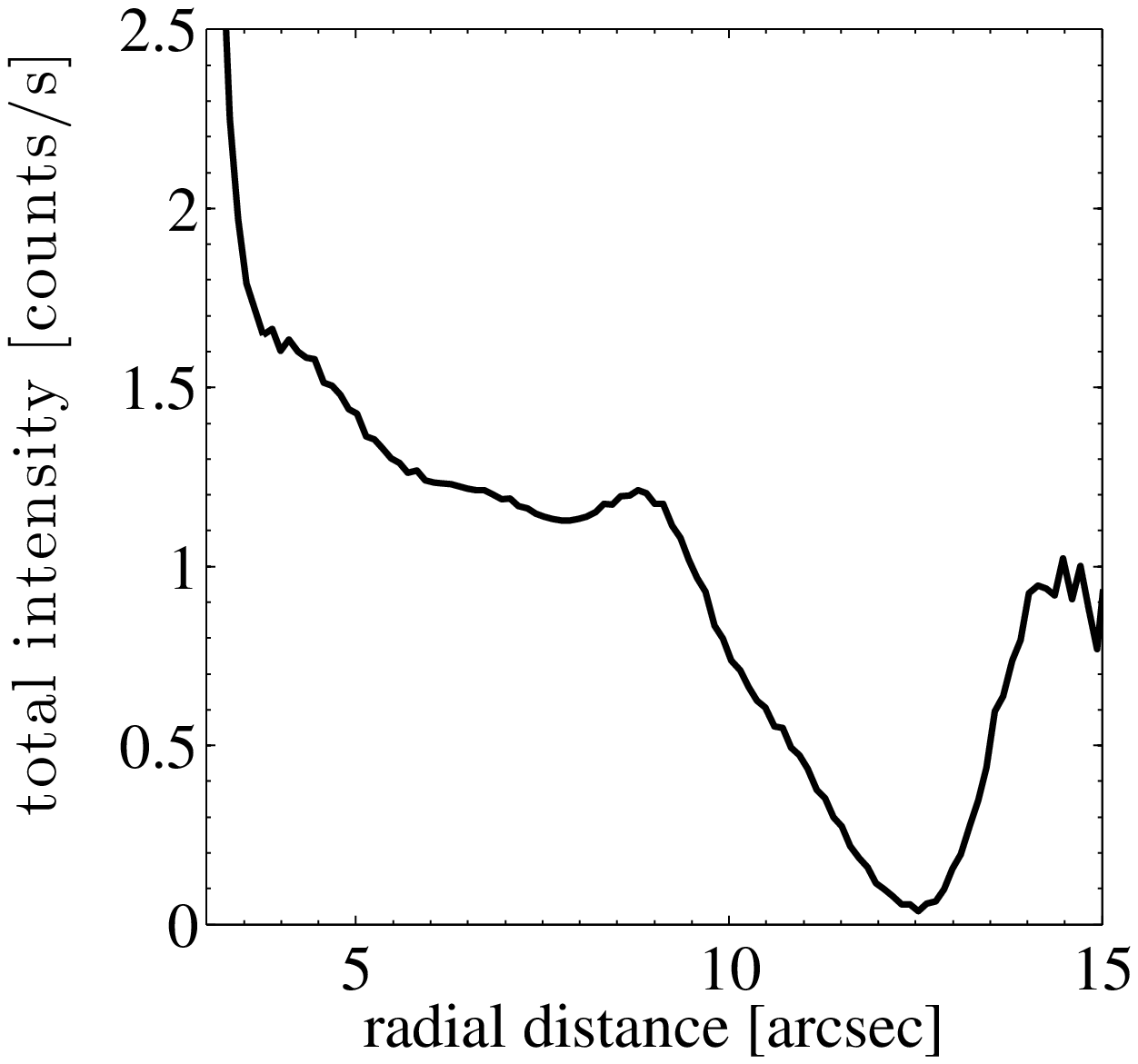}
\includegraphics[width=4.5cm]{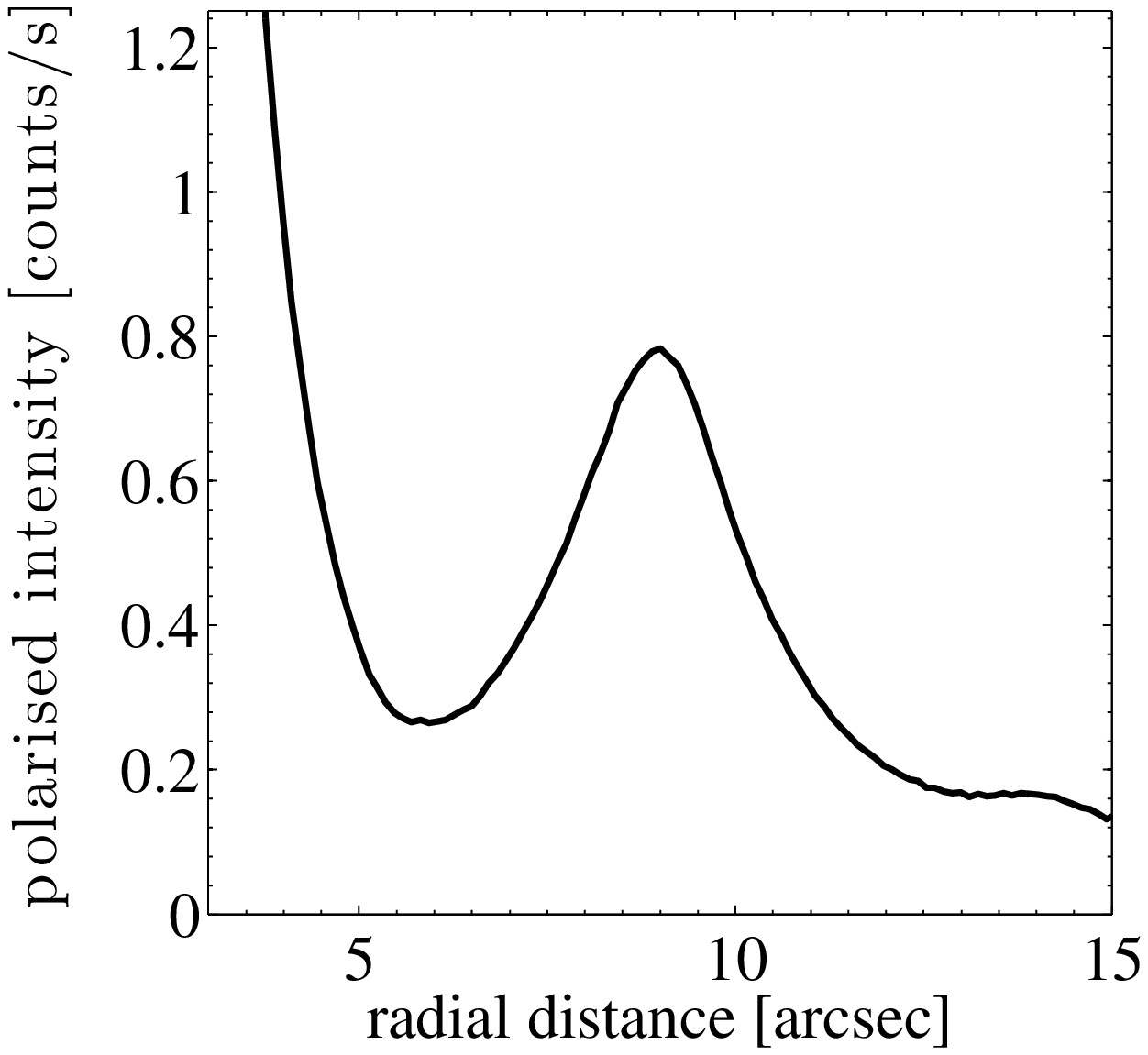}
\includegraphics[width=4.5cm]{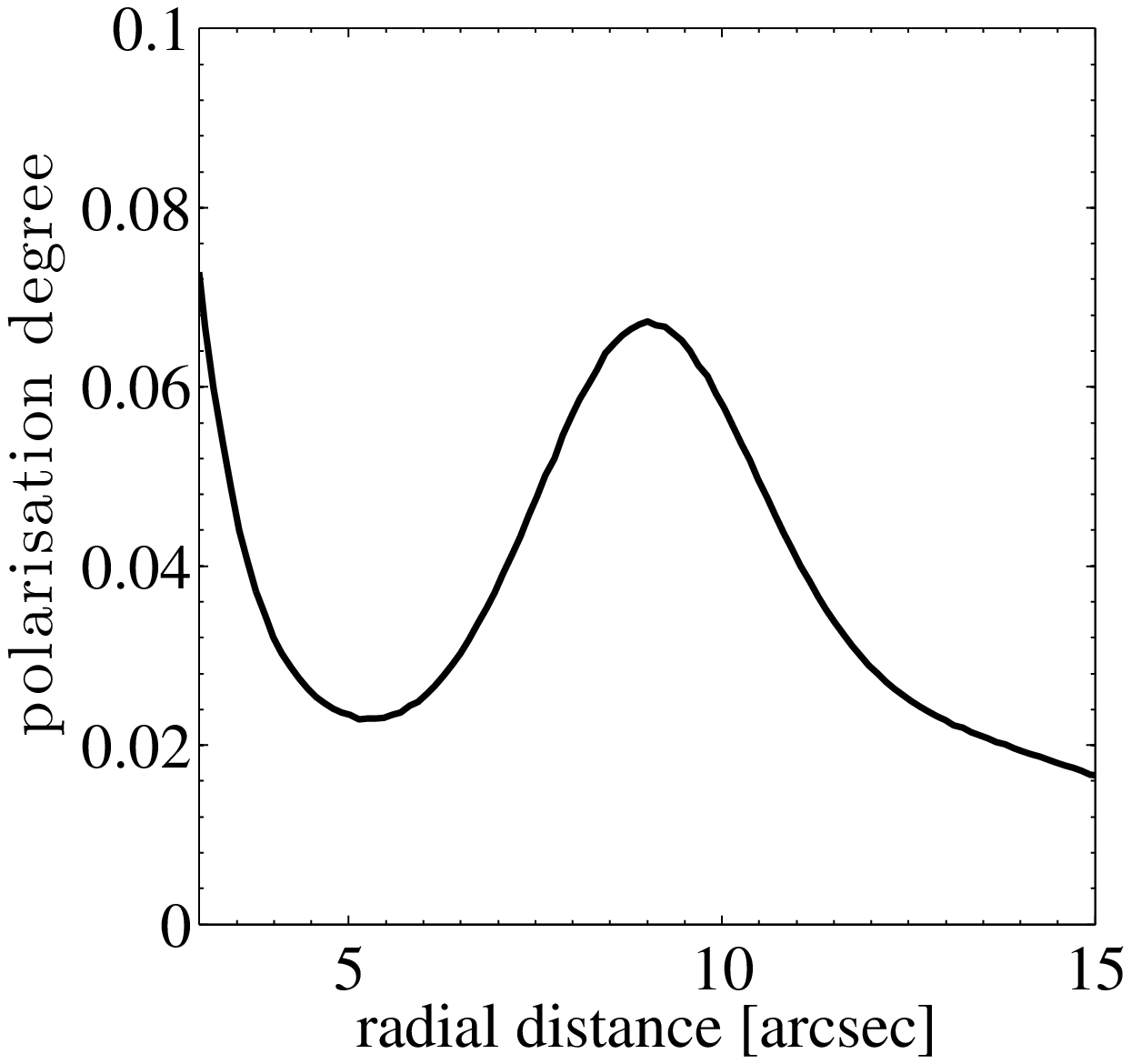}\\
\caption{Same as Fig.~\ref{f:v644RbandOct24} but in the V-band. The images are smoothed by a Gaussian kernel with a FWHM of four pixels.}
\label{f:v644VbandOct24}
\end{figure*}

Figures~\ref{f:rsclRbandOct19} to~\ref{f:v644VbandOct24} show the PolCor observations for R~Scl and V644~Sco in the R-band and V-band. For both sources the images of the observed total intensity, the total intensity after psf subtraction, the polarised intensity, and the lower limit of the polarisation degree are shown. For each of these we also present azimuthally averaged radial profiles (AARP). The profiles are averaged over regions that avoid artefacts introduced by the telescope spider. The detached shells are detected in the images of polarised emission at signal-to-noise ratios of S/N=$10-20$. Due to the difficulty of the template star subtraction (Sect.~\ref{s:psfsub}), only a lower limit of the polarisation degree is estimated. Table~\ref{t:shellresults} shows the determined shell radii and widths (defined as FWHM of a Gaussian distribution) for R~Scl and V644~Sco. 

\begin{figure}[h]
\centering
\includegraphics[width=7.5cm]{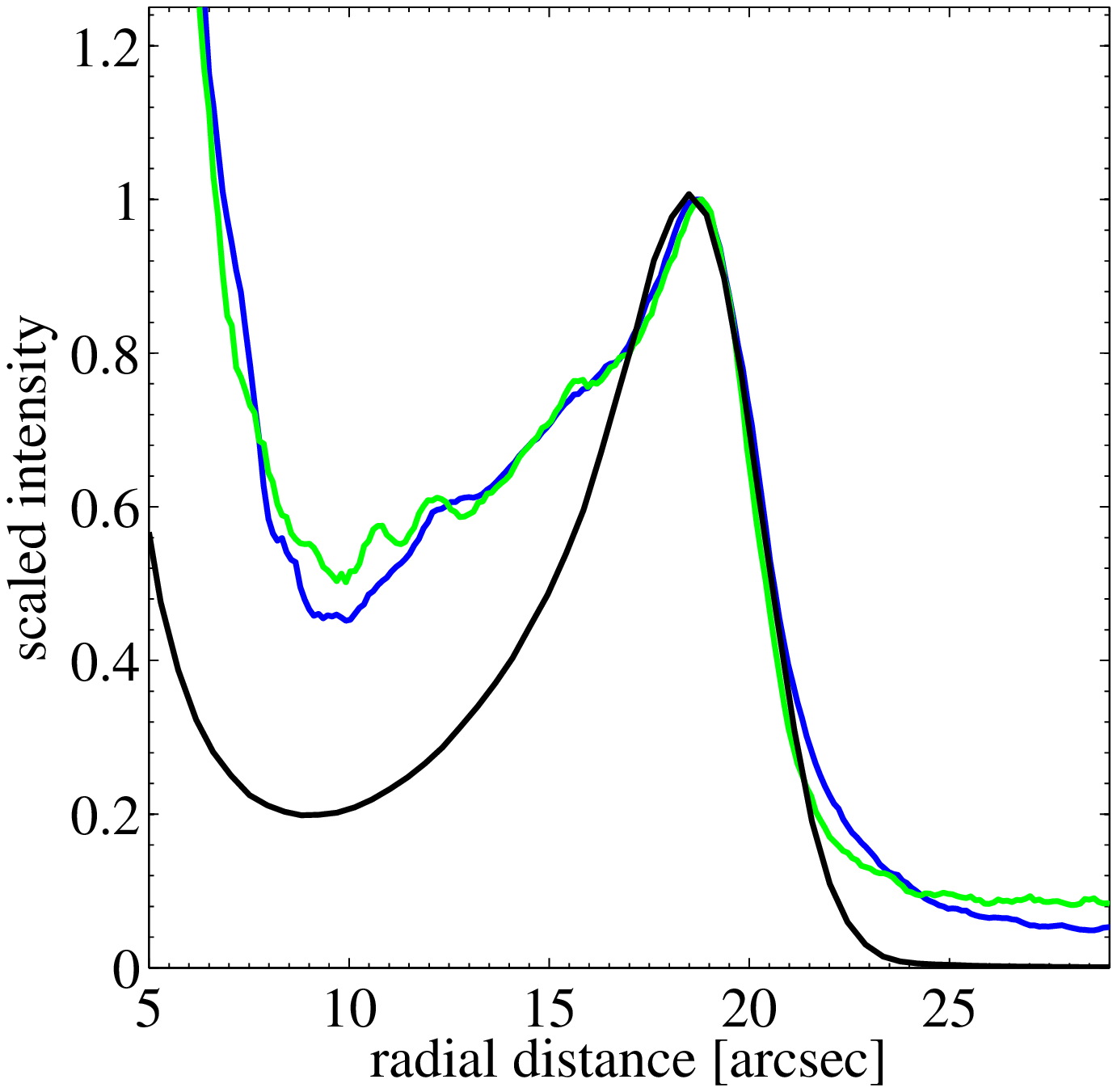}
\includegraphics[width=7.5cm]{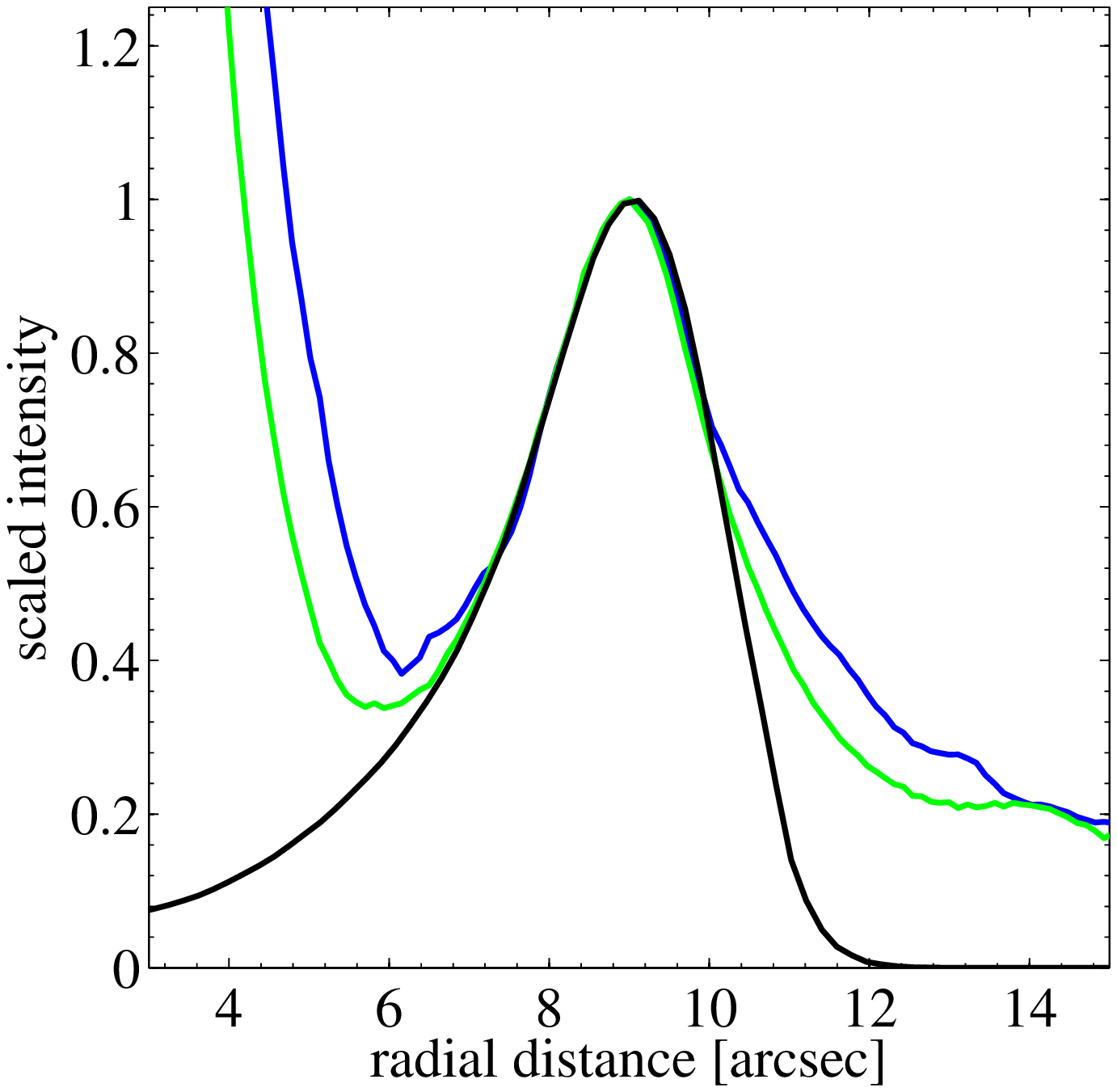}
\caption{The radial profiles of the polarised intensity compared to the radiative transfer models for R~Scl (top) and V644~Sco (bottom). The R-band and V-band data are plotted as blue and green lines, respectively, the model profiles as black lines. The profiles are normalised to 1. }
\label{f:csePmodel}
\end{figure}

The sizes and widths of the shells are determined by creating AARPs of the polarised intensity in the \emph{radmc-3d} model images. The peak of the intensity at the position of the shell in the model AARP is then scaled to the peak in the observed AARP at the position of the shell. The radius of the shell and the width are varied until a satisfactory fit to the observed profile is achieved (Fig.~\ref{f:csePmodel}). The radius is constrained by fitting the position of the peak, while the width is determined by fitting the model AARP to the \emph{outer} edge of the detached shell in the observed AARP. Both the radii and widths can be determined to within $\pm$0\farcs5.

In principle, the increase in intensity towards the inner regions could be used to constrain the density contrast between the present-day mass-loss rate and the detached shells. However, effects due to the coronographic mask and a potential polarisation of the central star make any such estimate extremely uncertain. 

\begin{table}
\caption{Measured shell radii (R) and shell widths ($\Delta$R -- FWHM of Gaussian distribution) for R~Scl and V644~Sco. Both can be determined to with $\pm$0\farcs5. The results from this paper are derived from radiative transfer models constrained by the polarised intensity images from PolCor (see Fig.~\ref{f:csePmodel}). The results are compared to previous estimates of the detached shells.}
\label{t:shellresults}
\centering
\begin{tabular}{l c c c c c l}
\hline\hline
		& \multicolumn{2}{c}{R~Scl}	&  &\multicolumn{2}{c}{V644~Sco} &\\
		&	R[\arcsec]	& $\Delta$R[\arcsec]		& & R[\arcsec]	& $\Delta$R[\arcsec] &\\
\hline
\textbf{this paper}	& 19.7	& 3.2			& & 9.4		& 2.0&\\
GD2001\tablefootmark{a}		& 20.7	& --			& & --		& -- &\\
GD2003$^b$		& 20.0	& 2.0			& & --		& -- &\\
S2005 (dust)$^c$	& 27.0	& --			& & 17.0	& -- &\\
S2005 (gas)$^c$	& 20.1	& --			& & 10.5	& -- &\\
O2010$^d$		& 19.2	& 1.2			& & --		& -- &\\
M2012$^e$ 		& 18.5	& 1.3			& & --		& -- &\\
\hline\hline
\multicolumn{7}{l}{\tablefoottext{a}{based on observations stellar light scattered in NaD and K }}\\
\multicolumn{7}{l}{resonance lines}\\
\multicolumn{7}{l}{\tablefoottext{b}{based on polarised, dust-scattered stellar light}}\\
\multicolumn{7}{l}{\tablefoottext{c}{based on dust and gas radiative transfer models}}\\
\multicolumn{7}{l}{\tablefoottext{d}{based on observations of dust scattered stellar light}}\\
\multicolumn{7}{l}{\tablefoottext{e}{based on fit to observed CO($3-2$) line emission}}\\
\multicolumn{7}{l}{ with ALMA}\\
\end{tabular}
\end{table}

\subsection{R~Scl}
\label{s:rsclresults}

A shell radius for R~Scl is determined from the dust-scattered polarised light images to 19\farcs7 and a FWHM width of the Gaussian shell to 3\farcs2. The polarisation degree reaches values of $>$20\% at the position of the shell, showing that the dust is indeed located in a thin shell around the star. A comparison with determined shell radii and widths using different probes for the dust and gas is presented in Table~\ref{t:shellresults}. Our results are consistent with the radius derived by (GD2003). The determined radius also fits well with the radius determined in the HST images in O2010. We derive a larger FWHM for the shell than O2010. However, the HST images only cover $\approx1/3$ of the total shell. The region imaged in the HST data is dominated by a bright arc along the limb-brightened shell, while the \emph{radmc-3d} models in this work are fit to the AARP of the entire shell, which may lead to a broader average width. In particular the apparent flattening of the Southern part of the shell will lead to a slightly broader shell. The shell radius derived in the dust models in S2005 has a large uncertainty ($\pm14$\arcsec). These models depend critically on the assumed grain properties. If the grains are too small for example, the radius of the shell will be overestimated, since the grains have to be moved to larger distances to be cool enough to achieve the right temperature and give the correct infrared flux. 

Imaging of the detached gas shell around R~Scl in stellar light scattered in resonance lines of NaD and K~(GD2001) gives results that are consistent with ours. The radius of the shell in the CO data from ALMA (M2012) is determined by measuring the peak of the emission in the AARP. This may lead to underestimating the size, due to limb-brightening along the inner edge of the shell, slightly moving the observed peak inwards compared to the density peak of the shell. This was already observed in M2010 and O2010. The width of the shell determined in ALMA data is limited by the shell being marginally resolved (with a beam-size of $\approx$1\farcs2). Hence, the results presented here for the radius and the width of the dust shell are consistent with the measurements of the CO($3-2$) emission line in the ALMA data.

\begin{figure*}
\centering
\includegraphics[width=8cm]{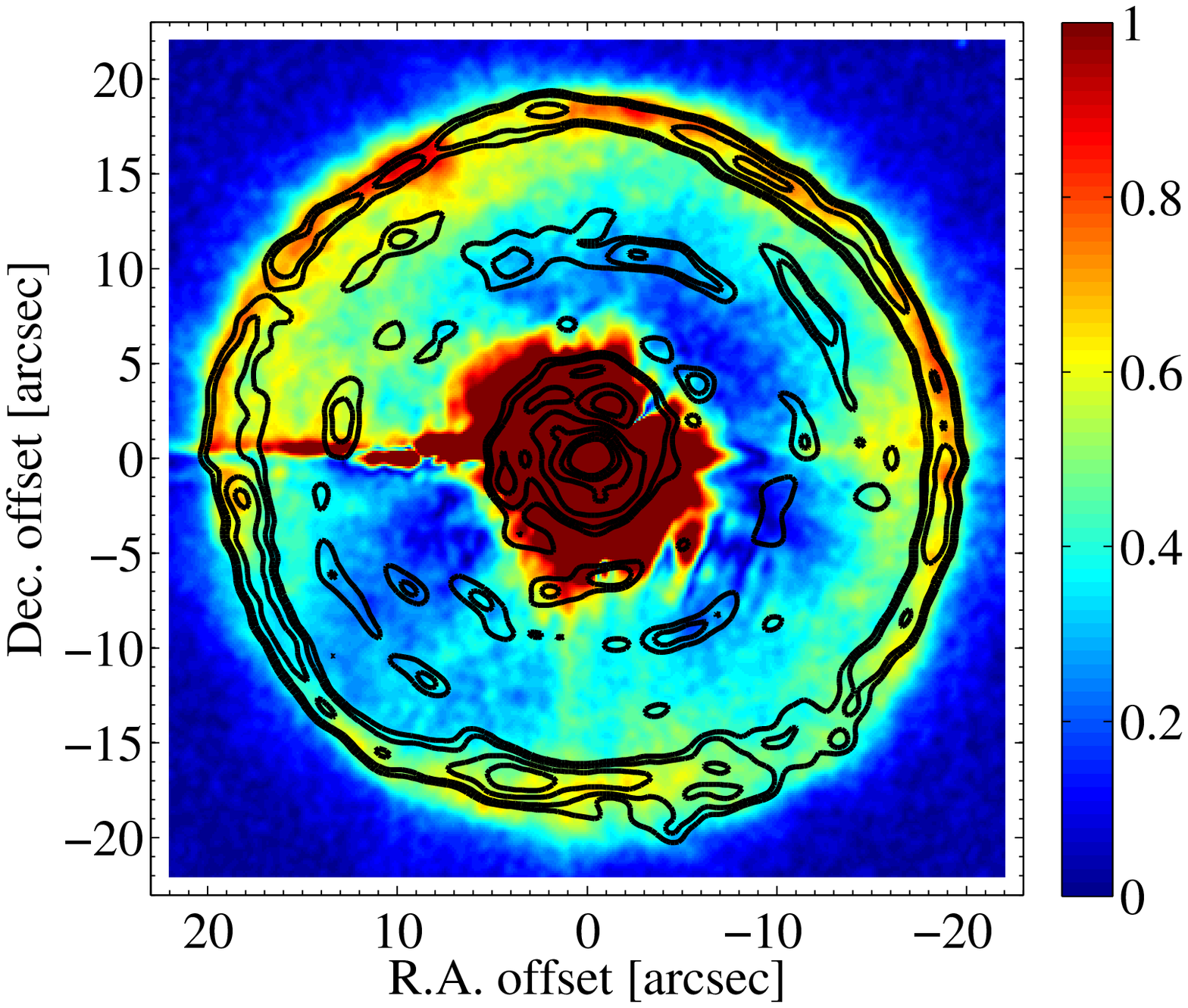}
\includegraphics[width=8cm]{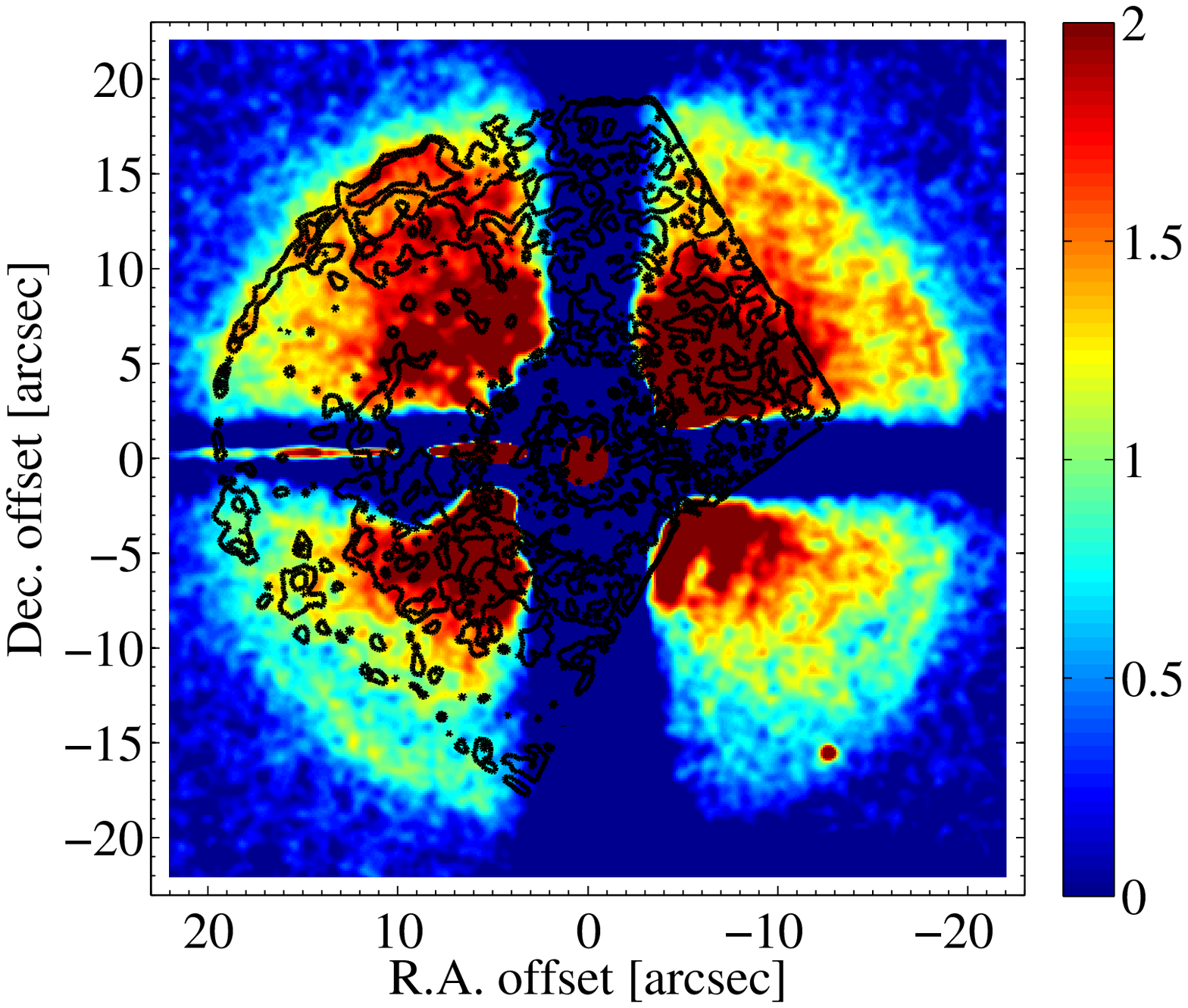}
\caption{Comparison of the PolCor polarised intensity image for R~Scl in the R-band with the ALMA band 7 data at the stellar $v_{lsr}$ (black contours, left) and the PolCor total intensity image with the Hubble Space Telescope data at 0.8$\mu$m (black contours, right). The colour scale shows the PolCor images in counts/s. }
\label{f:csePcomp}
\end{figure*}

In Fig.~\ref{f:csePcomp} (left) we compare the polarised intensity PolCor image in the R-band with the ALMA image of the CO($3-2$) emission centred on the $v_{lsr}$ in a 0.5\,\kms\,bin. The image of the polarised intensity shows the morphology of the dust in the detached shell in the plane of the sky, while the image at the stellar $v_{lsr}$ from ALMA shows the CO($3-2$) emission in the plane of the sky. Assuming that the gas emission (roughly) traces the density distribution of the gas, the distribution of the dust and gas can be compared. The ALMA contours trace the PolCor image closely, showing that the spatial distribution of the dust and gas coincides. In particular, the flattening of the shell in the south is present in both images, as well as an apparent gap in the south-western part of the shell. The peaks of the emission also generally coincide between the dust and the gas (Fig.~\ref{f:parad}), although there are a few exceptions, most notably the emission peak in the molecular gas at a position angle P.A.=120$^{\circ}$ where the dust has a clear gap. Nevertheless, the comparison between the PolCor and ALMA data suggests that the distribution of the dust and gas, including different clumps, is the same.

The total intensity PolCor image on the other hand should be directly comparable to the HST image observed by O2010. Figure~\ref{f:csePcomp} (right) shows the contours of the HST image in the f814 filter on top of the PolCor R-band total intensity image where the stellar psf has been subtracted. The psf subtraction in the HST data are of much higher quality, and structures observed close to the star are more reliable than for the PolCor data. The comparison between the two images shows that in particular for the shell the PolCor image shows the same features as the HST observations. The advantage of the PolCor data is that it covers the entire shell, while the limited field of view in the HST data misses potentially interesting features, such as the flattening of the shell in the southern part and an apparent gap to the south-west.

Except for the uncertain estimate from the dust modelling in S2005, all estimates of the size and width of the detached shell of dust and gas around R~Scl give consistent results. The average value for the radius of the shell (taking all measurements into account) is 19\farcs5$\pm$0\farcs7. The width of the shell varies more. This however depends on the type of observation and measurement method. An average width of $\approx2$\arcsec, with 1\arcsec and 3\arcsec as lower and upper limits for a varying width, appears reasonable. The observations of the detached shell presented here are the only images with high-spatial resolution of the \emph{entire} detached shell of dust around R~Scl, allowing us to measure the size of the shell with high accuracy.

\subsection{V644~Sco}
\label{s:v644results}

V644~Sco was observed in CO emission for the first time in~\cite{olofssonetal1996}, showing the typical double-peaked line profile indicative of an expanding detached shell. They mapped the shell in the CO($3-2$) transition in a 49-point map with a 16\arcsec\, beam and a 7\farcs5 spacing. Although the emission appears to be partially resolved at velocities close to the v$_{lsr}$, they conclude that the detached shell must have a radius of less that 7\arcsec. S2005 determine the detached shell to have a radius of 10\farcs5 (Table~\ref{t:shellresults}) based on molecular line modelling, and a radius of 17\farcs0$\pm$7\arcsec\, based on the dust modelling. Also here the dust modelling suffers from large uncertainties. In the PolCor data (Figs.~\ref{f:v644RbandOct24} and~\ref{f:v644VbandOct24}) the shell around V644~Sco can clearly be seen at a radius of 9\farcs4. The width of a Gaussian shell fit to the AARP is 2\farcs0. This is the first time the radius and width of the shell around V644~Sco have been measured \emph{directly}, instead of being inferred from radiative transfer modelling. The radius derived from the molecular line modelling is remarkably close to the observed value in the PolCor data, lending strength to the results of the CO models in S2005. 

The polarised data clearly show the detached shell, as well as structure/inhomogeneities within the shell. As with similar observations of other detached shell sources, V644~Sco shows a remarkable overall spherical symmetry. We only determine a lower limit of the polarisation degree due to the difficulties with the psf subtraction (Sect.~\ref{s:psfsub}). However, in all images the shell clearly appears with the degree of polarisation peaking at the position of the shell seen in the polarised images. As with R~Scl, this implies that the dust is distributed in a thin shell with the determined radius and width. This indicates that the shell originated in a spherical symmetric mass loss event from the star, and evolved as an expanding density enhancement. The density enhancement may be retained through interaction with a surrounding spherically symmetric medium as described by \cite{mattssonetal2007}. 

\section{Discussion}
\label{s:discussion}

\subsection{The detached shells around R~Scl and V644~Sco}
\label{s:discshells}

\begin{figure*}
\centering
\includegraphics[width=16cm]{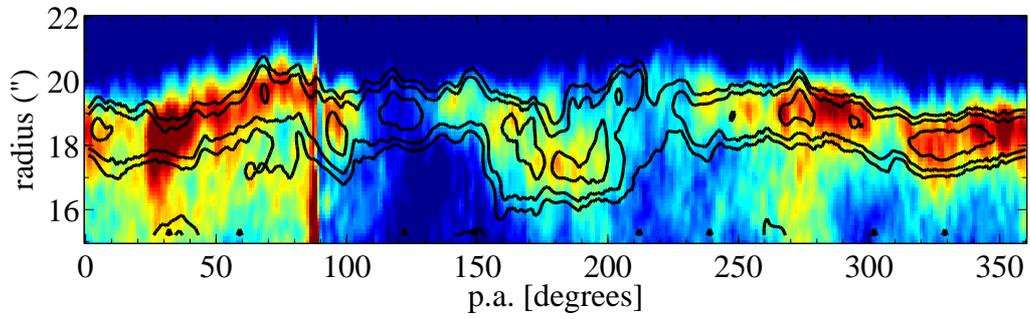}
\caption{Position angle (p.a.) vs. radius in the range from 15\arcsec to 11\arcsec showing the detached shell around R~Scl. The colour image shows the polarised intensity from PolCor in the R-band, the contours the ALMA CO($3-2$) emission at the stellar $v_{lsr}$. The p.a. start north and moves counter-clockwise. }
\label{f:parad}
\end{figure*}

The PolCor images of R~Scl allow for a direct comparison with the ALMA observations of the CO($3-2$) line emission (Fig.~\ref{f:csePcomp}, left, and Fig.~\ref{f:parad}). The good correlation between the PolCor data and the HST and ALMA observations confirms that the observations reliably measure the radius and width of the observed detached shells. These are the most detailed observations of the shell of dust around R~Scl to date. For V644~Sco we for the first time set reliable and direct constraints on the size and width of the detached shell of dust.

For U~Ant, \cite{maerckeretal2010} find that the dust has (mostly) separated from the gas, with a detached shell dominated by dust outside a detached shell dominated by gas. This implies a drift velocity between the dust and gas, as is expected in a dust-driven wind. However, the comparison between the PolCor data and ALMA data for R~Scl show a remarkable correlation between the dust and gas. The overall shape of the shell is reproduced in both observations, including the flatter southern part of the shell, while even individual clumps of dust and gas coincide. This implies a common evolution of the dust and gas around R~Scl. Also for V644~Sco, U~Cam, and DR~Ser \citep[O2010;][]{ramstedtetal2011} the derived sizes for the detached dust shells coincide with the sizes determined for the gaseous detached shell. The shells around R~Scl, V644~Sco, U~Cam, and DR~Ser are 1000-2000 years younger than the shell around U~Ant (S2005). \cite{maerckeretal2010} suggest a common origin of the dust and gas in the shells around U~Ant during a thermal pulse, with a drift velocity of a few \kms. \cite{kerschbaumetal2010} only detect the inner detached shell in thermal dust emission in far-infrared PACS observations, while \cite{arimatsuetal2011} detect both shells in mid-infrared observations. This implies the presence of small, hot grains in the inner shell, and larger grains in the outer shell which have drifted away from the gas, confirming the conclusions by \cite{maerckeretal2010}. It is possible that the dust has not yet separated significantly from the younger detached shell sources. The separation of the dust and gas may also be affected by the surrounding medium, and hence the pre-pulse mass-loss rate. 

The observed structures support the hypothesis in which the detached shells originate in the high mass-loss-rate phase during a thermal pulse. The shells then expand away from the star, while the declining post-pulse mass-loss rate causes the shell to appear as detached. Comparing the results of dust-scattered light one can conclude that the radii of the shells increase as a function of age as expected. Further, with the exception of U~Cam, the widths of the shells are similar ($\approx1000-2000$AU -- note that while the angular sizes of the shells can be relatively well determined, the physical sizes are very dependent on the estimated distance to the sources). This is consistent with the creation of a detached shell during a mass-loss eruption and a subsequent interaction with a surrounding medium, keeping the absolute width of the shell constant \citep{steffenco2000}.

\subsection{The importance of constraining the shells}
\label{s:importance}

Assuming distances of 290 pc and 700 pc to R~Scl and V644~Sco, respectively, we obtain physical radii of the detached shells of 8.4$\times10^{16}$\,cm and 9.8$\times10^{16}$\,cm, respectively. The widths of the shells are 8.7$\times10^{16}$\,cm and 2.1$\times10^{16}$\,cm for R~Scl and V644~Sco, respectively. Constraining the physical sizes of the shells is important to constrain hydrodynamical models. Together with the velocity information in the molecular data, and information on the spatial correlation of the dust and gas, the age of the shell can be constrained. The radii of the shells depend on the age of the shell and the expansion velocity as a function of time. The width of the shell depends on the formation conditions during the thermal pulse, and the subsequent interaction with the circumstellar medium \citep{steffenco2000,mattssonetal2007}. Hydrodynamical models have to model the pre-pulse mass loss, the formation of the detached shell during a thermal pulse, and the subsequent evolution of the shell. The models are ultimately constrained by matching the observed parameters of the shells. The presented observations hence constrain the timescales and conditions of the pre-pulse, pulse, and post-pulse evolution (M2012).

The total mass in the detached shell likewise depends on the formation process and subsequent evolution of the shell. However, accurate estimates of the dust masses require high-quality observations at optical wavelengths in different bands (to constrain the grain size distribution of small grains), as well as observations at far-infrared and radio wavelengths to also constrain the population of large dust grains. The shapes and composition of the grains are also large uncertainties, and have to be constrained by spectral observations. For some detached shell objects far-infrared observations exist~\citep{kerschbaumetal2010,izumiuraetal2011,mecinaetal2014}. R~Scl has been observed with the Large Bolometer Camera (LABOCA) on the Atacama Pathfinder Experiment (APEX) at 850\,$\mu$m, and we are currently performing observations with LABOCA of three additional detached shell sources (V644~Sco, U~Ant, and DR~Ser) to probe the contribution of large dust grains to the total mass. 

The drawback of observations in the far-infrared and radio is that the spatial resolution is relatively low. The optical observations in polarised light obtained here can be used to effectively constrain the size and width of the detached shells in the modelling of the thermal dust emission, leading to more accurate dust-mass estimates. In particular, a comparison to the PACS \citep{coxetal2012} and LABOCA observations of R~Scl will determine the properties of the detached dust shell with similar detail as the observations of the detached gas shell.

\section{Conclusions}
\label{s:conclusions}

The observations of polarised, dust-scattered stellar light presented here clearly show the detached dust shells around the carbon AGB stars R~Scl and V644~Sco. The data presented here are the most detailed images of the entire dust shell around R~Scl to date. They allow us to measure the size and width of the detached shell of dust and determine the morphology of the shell with high precision. For V644~Sco this is the first time that the detached shell of dust was constrained directly. 

For R~Scl a direct comparison of the dust-scattered optical emission observed here with ALMA observations of CO($3-2$) (M2012) and optical Hubble Space Telescope images (O2010) shows that the dust and gas in the detached shell coincide spatially to a high degree. This implies that the dust and gas have evolved together since the detached shell was created. 

For V644~Sco this is the first time that the radius and width of the detached dust shell were measured directly. Comparison with radiative transfer models of the molecular emission (S2005) indicate also here a close correlation between the dust and gas. 

The creation of detached shells of dust and gas by thermal pulses has been described previously, both observationally and theoretically \citep[e.g.][]{schoieretal2005,mattssonetal2007}. The results found here are in line with a scenario where the increase in mass-loss rate and expansion velocity during a thermal pulse leads to the formation of a detached shell of dust and gas. The derived radii and widths of the shells will help constrain more detailed hydrodynamical and radiative transfer models of the detached shells, and hence help constrain the formation and evolution of the shells since formation during the thermal pulse. 

The results demonstrate that observations of polarised, dust-scattered stellar light can trace intricate details in the circumstellar envelopes around AGB stars. They are particularly useful to constrain the structures in the detached shells. These observations were part of a pilot sample, and hence only have a limited amount of time spent per object. Already they give new constraints on the detached shells around R~Scl and, in particular,  V644~Sco. Dedicated observations with PolCor and/or similar instruments of the known detached shell sources will help constrain the evolution of the star during and after a thermal pulse. With ALMA, correlating the evolution of the dust with the evolution of the molecular gas will give the best observational description of the thermal pulse phenomenon to date.

\begin{acknowledgements}
M.L.L.-F. was supported by the Deutscher Akademischer Austausch Dienst (DAAD). The authors would like to thank W.H.T Vlemmings, E. De Beck, and H. Olofsson for their contribution in discussing the results.
\end{acknowledgements}

\bibliographystyle{aa} 
\bibliography{maercker.bib}

\begin{thebibliography}{30}
\expandafter\ifx\csname natexlab\endcsname\relax\def\natexlab#1{#1}\fi

\bibitem[{{Arimatsu} {et~al.}(2011){Arimatsu}, {Izumiura}, {Ueta}, {Yamamura},
  \& {Onaka}}]{arimatsuetal2011}
{Arimatsu}, K., {Izumiura}, H., {Ueta}, T., {Yamamura}, I., \& {Onaka}, T.
  2011, ApJ, 729, L19

\bibitem[{{Castro-Carrizo} {et~al.}(2010){Castro-Carrizo}, {Quintana-Lacaci},
  {Neri}, {Bujarrabal}, {Sch{\"o}ier}, {Winters}, {Olofsson}, {Lindqvist},
  {Alcolea}, {Lucas}, \& {Grewing}}]{castrocarrizoetal2010}
{Castro-Carrizo}, A., {Quintana-Lacaci}, G., {Neri}, R., {et~al.} 2010, A\&A,
  523, A59

\bibitem[{{Cox} {et~al.}(2012){Cox}, {Kerschbaum}, {van Marle}, {Decin},
  {Ladjal}, {Mayer}, {Groenewegen}, {van Eck}, {Royer}, {Ottensamer}, {Ueta},
  {Jorissen}, {Mecina}, {Meliani}, {Luntzer}, {Blommaert}, {Posch},
  {Vandenbussche}, \& {Waelkens}}]{coxetal2012}
{Cox}, N.~L.~J., {Kerschbaum}, F., {van Marle}, A.-J., {et~al.} 2012, A\&A,
  537, A35

\bibitem[{{Decin} {et~al.}(2012){Decin}, {Cox}, {Royer}, {Van Marle},
  {Vandenbussche}, {Ladjal}, {Kerschbaum}, {Ottensamer}, {Barlow}, {Blommaert},
  {Gomez}, {Groenewegen}, {Lim}, {Swinyard}, {Waelkens}, \&
  {Tielens}}]{decinetal2012}
{Decin}, L., {Cox}, N.~L.~J., {Royer}, P., {et~al.} 2012, A\&A, 548, A113

\bibitem[{{Dullemond}(2012)}]{dullemond2012}
{Dullemond}, C.~P. 2012, {RADMC-3D: A multi-purpose radiative transfer tool},
  astrophysics Source Code Library

\bibitem[{{Gonz{\'a}lez Delgado} {et~al.}(2001){Gonz{\'a}lez Delgado},
  {Olofsson}, {Schwarz}, {Eriksson}, \& {Gustafsson}}]{delgadoetal2001}
{Gonz{\'a}lez Delgado}, D., {Olofsson}, H., {Schwarz}, H.~E., {Eriksson}, K.,
  \& {Gustafsson}, B. 2001, A\&A, 372, 885

\bibitem[{{Gonz{\'a}lez Delgado} {et~al.}(2003){Gonz{\'a}lez Delgado},
  {Olofsson}, {Schwarz}, {Eriksson}, {Gustafsson}, \&
  {Gledhill}}]{delgadoetal2003a}
{Gonz{\'a}lez Delgado}, D., {Olofsson}, H., {Schwarz}, H.~E., {et~al.} 2003,
  A\&A, 399, 1021

\bibitem[{{Izumiura} {et~al.}(2011){Izumiura}, {Ueta}, {Yamamura}, {Matsunaga},
  {Ita}, {Matsuura}, {Nakada}, {Fukushi}, {Mito}, {Tanab{\'e}}, \&
  {Hashimoto}}]{izumiuraetal2011}
{Izumiura}, H., {Ueta}, T., {Yamamura}, I., {et~al.} 2011, A\&A, 528, A29

\bibitem[{{Karakas} \& {Lattanzio}(2007)}]{karakasco2007}
{Karakas}, A. \& {Lattanzio}, J.~C. 2007, PASA, 24, 103

\bibitem[{{Kerschbaum} {et~al.}(2010){Kerschbaum}, {Ladjal}, {Ottensamer},
  {Groenewegen}, {Mecina}, {Blommaert}, {Baumann}, {Decin}, {Vandenbussche},
  {Waelkens}, {Posch}, {Huygen}, {De Meester}, {Regibo}, {Royer}, {Exter}, \&
  {Jean}}]{kerschbaumetal2010}
{Kerschbaum}, F., {Ladjal}, D., {Ottensamer}, R., {et~al.} 2010, A\&A, 518,
  L140

\bibitem[{{Kim} {et~al.}(2013){Kim}, {Hsieh}, {Liu}, \& {Taam}}]{kimetal2013}
{Kim}, H., {Hsieh}, I.-T., {Liu}, S.-Y., \& {Taam}, R.~E. 2013, ApJ, 776, 86

\bibitem[{{Kim} \& {Taam}(2012)}]{kimco2012}
{Kim}, H. \& {Taam}, R.~E. 2012, ApJ, 744, 136

\bibitem[{{Knapp} {et~al.}(2003){Knapp}, {Pourbaix}, {Platais}, \&
  {Jorissen}}]{knappetal2003}
{Knapp}, G.~R., {Pourbaix}, D., {Platais}, I., \& {Jorissen}, A. 2003, A\&A,
  403, 993

\bibitem[{{Maercker} {et~al.}(2012){Maercker}, {Mohamed}, {Vlemmings},
  {Ramstedt}, {Groenewegen}, {Humphreys}, {Kerschbaum}, {Lindqvist},
  {Olofsson}, {Paladini}, {Wittkowski}, {de Gregorio-Monsalvo}, \&
  {Nyman}}]{maerckeretal2012}
{Maercker}, M., {Mohamed}, S., {Vlemmings}, W.~H.~T., {et~al.} 2012, Nature,
  490, 232

\bibitem[{{Maercker} {et~al.}(2010){Maercker}, {Olofsson}, {Eriksson},
  {Gustafsson}, \& {Sch{\"o}ier}}]{maerckeretal2010}
{Maercker}, M., {Olofsson}, H., {Eriksson}, K., {Gustafsson}, B., \&
  {Sch{\"o}ier}, F.~L. 2010, A\&A, 511, A37+

\bibitem[{{Mastrodemos} \& {Morris}(1999)}]{mastrodemosco1999}
{Mastrodemos}, N. \& {Morris}, M. 1999, ApJj, 523, 357

\bibitem[{{Mattsson} {et~al.}(2007){Mattsson}, {H{\"o}fner}, \&
  {Herwig}}]{mattssonetal2007}
{Mattsson}, L., {H{\"o}fner}, S., \& {Herwig}, F. 2007, A\&A, 470, 339

\bibitem[{{Mauron} \& {Huggins}(2006)}]{mauronco2006}
{Mauron}, N. \& {Huggins}, P.~J. 2006, A\&A, 452, 257

\bibitem[{{Mayer} {et~al.}(2011){Mayer}, {Jorissen}, {Kerschbaum}, {Mohamed},
  {van Eck}, {Ottensamer}, {Blommaert}, {Decin}, {Groenewegen}, {Posch},
  {Vandenbussche}, \& {Waelkens}}]{mayeretal2011}
{Mayer}, A., {Jorissen}, A., {Kerschbaum}, F., {et~al.} 2011, A\&A, 531, L4

\bibitem[{{Me{\v c}ina} {et~al.}(2014){Me{\v c}ina}, {Kerschbaum},
  {Groenewegen}, {Ottensamer}, {Blommaert}, {Mayer}, {Decin}, {Luntzer},
  {Vandenbussche}, {Posch}, \& {Waelkens}}]{mecinaetal2014}
{Me{\v c}ina}, M., {Kerschbaum}, F., {Groenewegen}, M.~A.~T., {et~al.} 2014,
  A\&A, 566, A69

\bibitem[{{Olofsson} {et~al.}(1996){Olofsson}, {Bergman}, {Eriksson}, \&
  {Gustafsson}}]{olofssonetal1996}
{Olofsson}, H., {Bergman}, P., {Eriksson}, K., \& {Gustafsson}, B. 1996, A\&A,
  311, 587

\bibitem[{{Olofsson} {et~al.}(1990){Olofsson}, {Carlstrom}, {Eriksson},
  {Gustafsson}, \& {Willson}}]{olofssonetal1990}
{Olofsson}, H., {Carlstrom}, U., {Eriksson}, K., {Gustafsson}, B., \&
  {Willson}, L.~A. 1990, A\&A, 230, L13

\bibitem[{{Olofsson} {et~al.}(1993){Olofsson}, {Eriksson}, {Gustafsson}, \&
  {Carlstrom}}]{olofssonetal1993a}
{Olofsson}, H., {Eriksson}, K., {Gustafsson}, B., \& {Carlstrom}, U. 1993,
  ApJS, 87, 267

\bibitem[{{Olofsson} {et~al.}(2010){Olofsson}, {Maercker}, {Eriksson},
  {Gustafsson}, \& {Sch{\"o}ier}}]{olofssonetal2010}
{Olofsson}, H., {Maercker}, M., {Eriksson}, K., {Gustafsson}, B., \&
  {Sch{\"o}ier}, F. 2010, A\&A, 515, A27+

\bibitem[{{Ramstedt} {et~al.}(2011){Ramstedt}, {Maercker}, {Olofsson},
  {Olofsson}, \& {Sch{\"o}ier}}]{ramstedtetal2011}
{Ramstedt}, S., {Maercker}, M., {Olofsson}, G., {Olofsson}, H., \&
  {Sch{\"o}ier}, F.~L. 2011, A\&A, 531, A148

\bibitem[{{Sch{\"o}ier} {et~al.}(2005){Sch{\"o}ier}, {Lindqvist}, \&
  {Olofsson}}]{schoieretal2005}
{Sch{\"o}ier}, F.~L., {Lindqvist}, M., \& {Olofsson}, H. 2005, A\&A, 436, 633

\bibitem[{{Steffen} \& {Sch{\"o}nberner}(2000)}]{steffenco2000}
{Steffen}, M. \& {Sch{\"o}nberner}, D. 2000, A\&A, 357, 180

\bibitem[{{Steffen} {et~al.}(1998){Steffen}, {Szczerba}, \&
  {Schoenberner}}]{steffenetal1998}
{Steffen}, M., {Szczerba}, R., \& {Schoenberner}, D. 1998, A\&A, 337, 149

\bibitem[{{Suh}(2000)}]{suh2000}
{Suh}, K.-W. 2000, MNRAS, 315, 740

\bibitem[{{Zubko} \& {Laor}(2000)}]{zubkoco2000}
{Zubko}, V.~G. \& {Laor}, A. 2000, ApJS, 128, 245

\end{thebibliography}

\end{document}